\documentclass[preprint,12pt]{elsarticle}

\usepackage{amssymb}
\usepackage{amsmath}
\usepackage{amsfonts}
\usepackage{graphicx}
\usepackage{booktabs}
\usepackage{textcomp,nicefrac}
\usepackage{siunitx}  %
\usepackage{subfigure}
\usepackage{orcidlink}
\usepackage{multirow}
\usepackage{makecell}
\usepackage{lineno}
\biboptions{sort&compress}

\journal{NIMA}

\begin{document}

\begin{frontmatter}




\title{Development of Small-pitch, Ultra-thin 3D Silicon Sensors at USTC}

\author[ustc]{{Kuo Ma}\orcidlink{0009-0004-7076-0889}}
\author[ustc]{De Zhang}
\author[ustc]{Shengjia He}
\author[ustc]{Tian'ao Wang}
\author[ustc]{Han Li}
\author[ustc]{Yu Nie}
\author[nrfc]{Xiuxia Wang}
\author[nrfc]{Jinlan Peng}
\author[ustc]{Zheng Liang}
\author[sinano]{Xiang Li}
\author[sinano]{Wenhua Shi}
\author[sklftic,ime,ucas]{{Manwen Liu}}
\author[kek]{Chuan Liao}
\author[ludong]{Zheng Li}
\author[ustc]{Zebo Tang}
\author[ustc]{Yanwen Liu\corref{cor1}}
\ead{yanwen@ustc.edu.cn}

\affiliation[ustc]{organization={Department of Modern Physics, University of Science and Technology of China}, city={Hefei 230026}, country={China}}
\affiliation[nrfc]{organization={Center for Micro and Nanoscale Research and Fabrication, University of Science and Technology of China}, city={Hefei 230026}, country={China}}
\affiliation[sinano]{organization={Suzhou Institute of Nano-Tech and Nano-Bionics (SINANO), Chinese Academy of Sciences (CAS)}, city={Suzhou 215123}, country={China}}
\affiliation[sklftic]{organization={State Key Laboratory of Fabrication Technologies for Integrated Circuits, Chinese Academy of Sciences}, city={Beijing 100029}, country={China}}
\affiliation[ime]{organization={Institute of Microelectronics, Chinese Academy of Sciences}, city={Beijing 100029}, country={China}}
\affiliation[ucas]{orgazition={School of Integrated Circuits, University of Chinese Academy of Sciences}, city={Beijing 100049}, country={China}}
\affiliation[kek]{organization={International Center for Quantum-field Measurement Systems for Studies of the Universe and Particles(QUP), High Energy Accelerator Research Organization (KEK)}, country={Japan}}
\affiliation[ludong]{organization={College of Integrated Circuits, Ludong University}, city={Yantai 264025}, country={China}}
\cortext[cor1]{Corresponding author}

\begin{abstract}
We report on the development of 3D silicon sensors at the University of Science and Technology of China (USTC). The sensors involve columnar electrodes (5 \SI{}{\micro m} in diameter) of both doping types, etched from the same wafer side. The p$^+$ electrodes pass through the epitaxial layer, whereas the n$^+$ electrodes stop at a short distance from the opposite side of the epitaxial layer. Based on the current processing capacity, the sensors feature an ultra-thin active thickness (50 \SI{}{\micro m}) and a small pixel size of 50 \SI{}{\micro m} $\times$ 50 \SI{}{\micro m} or 25 \SI{}{\micro m} $\times$ 25 \SI{}{\micro m}. This R$\&$D project aims to establish a sensor technology to simultaneously measure position and time information at the single-pixel level. The first fabrication run, using a merged layout containing both pixel sizes, has been completed. The design and fabrication of the sensors, followed by their characterization, including Current–Voltage ($I-V$) measurements and $^{90}$Sr beta source measurements, are reported in this paper.
\end{abstract}




\end{frontmatter}


\section{Introduction}

Modern collider experiments in High Energy Physics (HEP) are required to deal with high collision rates. To facilitate track reconstruction and support particle identification, a novel tracking technique called 4D tracking has been proposed, which requires sensors with both good spatial and temporal resolution at the single-pixel level.

3D silicon detectors have been shown impressive radiation tolerance in HEP experiments. Take the ATLAS experiment at CERN as an example, the double-sided 3D sensors with 50 \SI{}{\micro m} $\times$ 250 \SI{}{\micro m} pixel size and 230 \SI{}{\micro m} active thickness were firstly used in the ATLAS Insertable B-Layer (IBL)~\cite{DAVIA2012321}, where the maximum irradiation fluence is 5~$\times$~10$^{15}$ n$_\mathrm{{eq}}$/cm$^2$ (1 MeV equivalent neutron per cm$^2$). At the High-Luminosity LHC (HL-LHC), the single-sided 3D sensors with a small size (25 \SI{}{\micro m} $\times$ 100 \SI{}{\micro m} or 50 \SI{}{\micro m} $\times$ 50 \SI{}{\micro m}) and thin active thickness (150 \SI{}{\micro m}) will be installed in the ATLAS inner tracker~\cite{10.3389/fphy.2021.624668} to withstand the expected maximum fluence of 1.2~$\times$~10$^{16}$ n$_\mathrm{{eq}}$/cm$^2$ and cope with occupancy in the innermost layer. 

In addition to the excellent radiation hardness, the signal response of the 3D silicon sensors is also very fast due to the short drift path of the carriers. Small cell 3D sensors could deliver a good time resolution at appropriate bias voltage, which can be compared with Low Gain Avalanche Detectors (LGADs)~\cite{Kramberger:2019ygz}. The test results show that ATLAS IBL 3D sensors with 50 \SI{}{\micro m} $\times$ 50 \SI{}{\micro m} pixel size can give a time resolution of 31 ps at 80 V and $-$20 $^{\circ}\mathrm{C}$~\cite{Diehl:2024kli}. Replacing columnar electrodes with trenched electrodes is a possible method to optimize the distribution of the electric field~\cite{5733383,LI201190}, which could improve the time resolution. This proposal has been adopted in the INFN TimeSPOT Project~\cite{MENDICINO201924,FORCOLIN2020164437,Brundu_2021}, which is dedicated to developing 3D trenched sensors for timing applications. The beam test results of 55 \SI{}{\micro m} $\times$ 55 \SI{}{\micro m} test structures with 150 \SI{}{\micro m} active thickness achieve an outstanding time resolution of the order of 10 ps at appropriate bias voltage and temperature, even after extreme radiation up to 1~$\times$~10$^{17}$ n$_\mathrm{{eq}}$/cm$^2$~\cite{10.3389/fphy.2024.1393019}.

The small-pitch 3D trenched sensor technology has shown even higher time resolution, but there are some limitations in its application for 4D tracking. The capacitance is larger than that of the small-pitch 3D columnar sensors, which impacts the noise. Due to the existence of highly doped electrodes in the vertical direction, the fill factor of the 3D sensor is not 100$\%$ for perpendicular tracks. While 3D sensors with columnar electrodes could be fully efficient for inclined tracks, the efficiency of 3D sensors with trenched electrodes is limited by the tracks at certain angles which don't pass through the active region. Thus, optimizing 3D sensors with columnar electrodes is still an interesting topic. 
3D columnar designs with a small pitch and different arrangements have been studied. The simulation results show that the intrinsic time resolution of the sensor with 45 \SI{}{\micro m} pitch and two electrodes (2E) could be in the lower part of the 10 ps to 45 ps range~\cite{Loi:2025aic}; without considering the contribution of electronics, the time resolution of the sensor with 25 \SI{}{\micro m} pitch and one electrode (1E) is approximately 13 ps at 100~V and \SI{-20}{\celsius}~\cite{Kramberger:2023qnh}. Recently, the gain effect has been observed of the 25 \SI{}{\micro m}-pitch sensor before irradiation~\cite{Gentry:2024jgn}. In addition, a novel 3D trench-column sensor with a pitch of \SI{35}{\micro m} and a narrow n$^+$ column was developed; gain was also observed in this sensor, although its time resolution still needs improvement~\cite{11271512,LIU2027172084}.

The goal of this work is to develop 3D column designs with two small pitches. While conducting research on currently common 50 \SI{}{\micro m} pitch, the study of a smaller pitch (25 \SI{}{\micro m} pitch) was also carried out. In order to explore a reliable processing procedure, the improved fabrication process was developed at USTC, including the deep reactive ion etching (DRIE) and thermal diffusion doping processes, and the first attempt was conducted on epitaxial wafers with 50 \SI{}{\micro m} active thickness. This paper is organized as follows: Section~\ref{Design and Simulations} describes the design and simulation of the 3D sensors. Section~\ref{Fabrication} details the layout designs and the fabrication process. Section~\ref{Measurement Setup} introduces the setup of Current-Voltage ($I-V$) measurement and $^{90}$Sr beta source measurement, as well as the design of the dedicated preamplifier board. Section~\ref{Measurement results} presents the results of the leakage current, breakdown voltage and time resolution. Section~\ref{conclusion} gives the conclusion and outlook.

\section{Design and simulation}
\label{Design and Simulations}

\begin{figure}[htbp]
    \centering
    \subfigure[]{\includegraphics[width=2.0in]{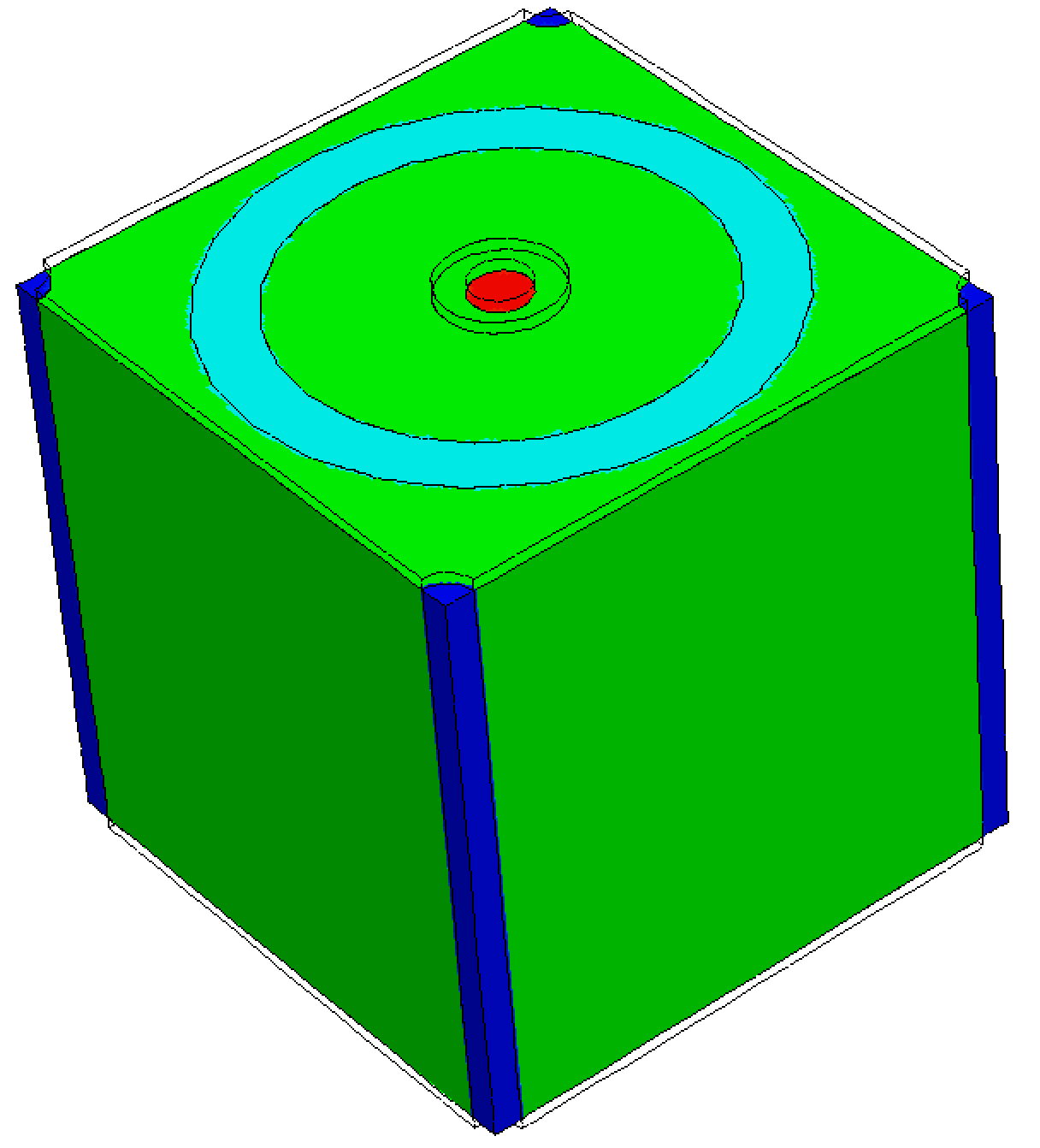}\label{p50_3d}}
    \subfigure[]{\raisebox{0.5cm}{\includegraphics[width=2.0in]{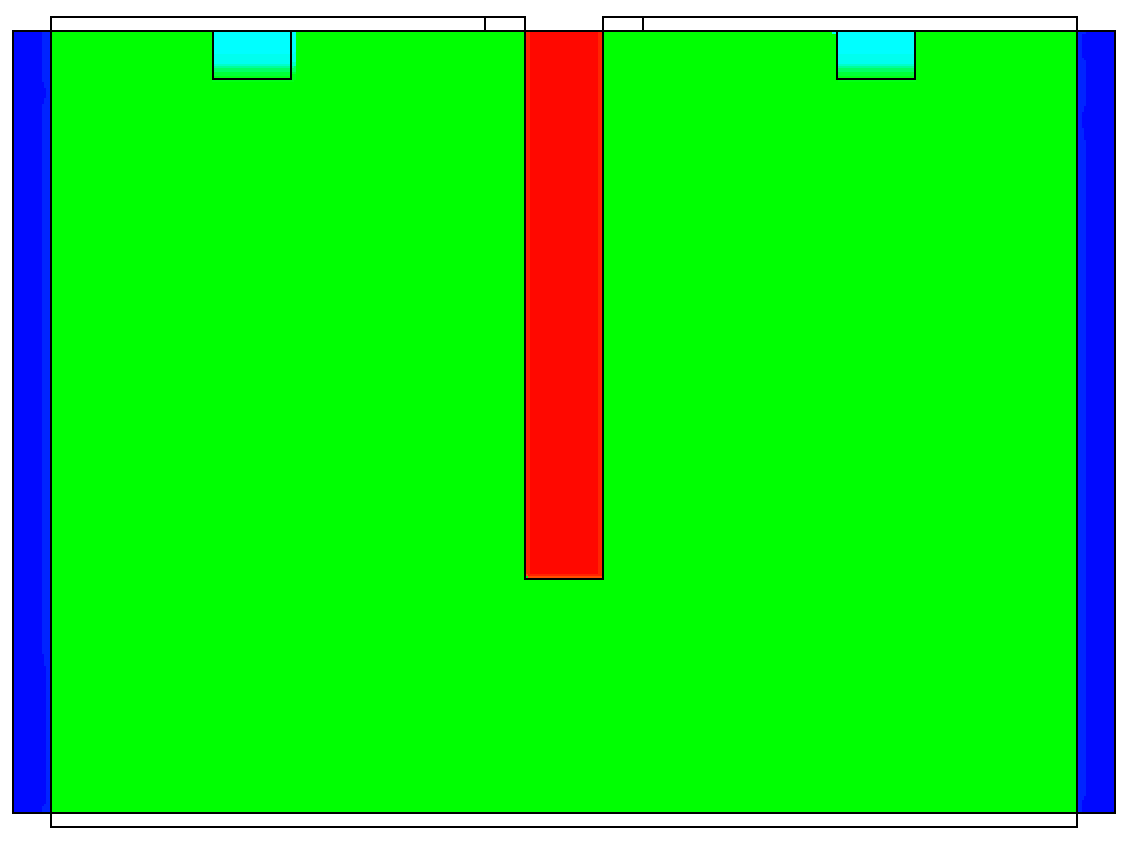}}}
    \subfigure{\raisebox{0.5cm}{\includegraphics[width=1.2in]{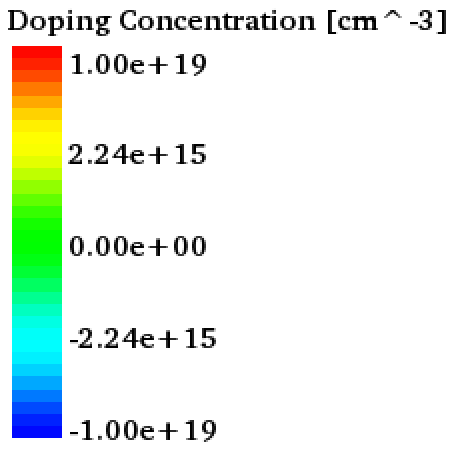}}}
    \caption{(a) 3D view of the sensor unit cell. (b) Diagonal cross section of the sensor unit cell.}
    \label{simulationcell}
\end{figure}

The simulations are carried out with TCAD simulation software~\cite{sentaurus_user_guide}. Fig.~\ref{simulationcell} shows the simulated unit, where a basic square cell consists of a central n$^+$ column and four p$^{+}$ columns located at the corners. The p$^+$ columns pass through the substrate, whereas the n$^+$ column reaches to a position 15 \SI{}{\micro m} from the backside. The diameter of the columns is 5 \SI{}{\micro m}. The doping concentration is 1~$\times$~10$^{19}$ cm$^{-3}$ with boron (B) for p$^+$ and phosphorus (P) for n$^+$, respectively. The thickness of the p-type epitaxial layer is 50 \SI{}{\micro m} and the B doping concentration is 1~$\times$~10$^{12}$ cm$^{-3}$. To isolate the n-type columns at the Si and SiO$_2$ interface~\cite{4337370}, a B-implanted p-stop structure is added (light blue region in Fig.~\ref{simulationcell}). The inner radius is 17.5 \SI{}{\micro m} and the width is 5 \SI{}{\micro m} for the structure with a pixel size of 50 \SI{}{\micro m} $\times$ 50 \SI{}{\micro m}, whereas the inner radius is 8 \SI{}{\micro m} and the width is 3 \SI{}{\micro m} for the structure with a pixel size of 25 \SI{}{\micro m} $\times$ 25 \SI{}{\micro m}. The physical models and the parameters used follow those reported in~\cite{BaselgaBacardit:2016lks}. The silicon dioxide surface charge is 4~$\times$~10$^{11}$~cm$^{-2}$. All simulations were conducted for the detector operating at a temperature of 20~$^\circ\mathrm{C}$ and the frequency used to simulate the capacitance was 10 kHz.

\begin{figure}[htbp]
    \centering
    \subfigure[]{\includegraphics[width=2.6in]{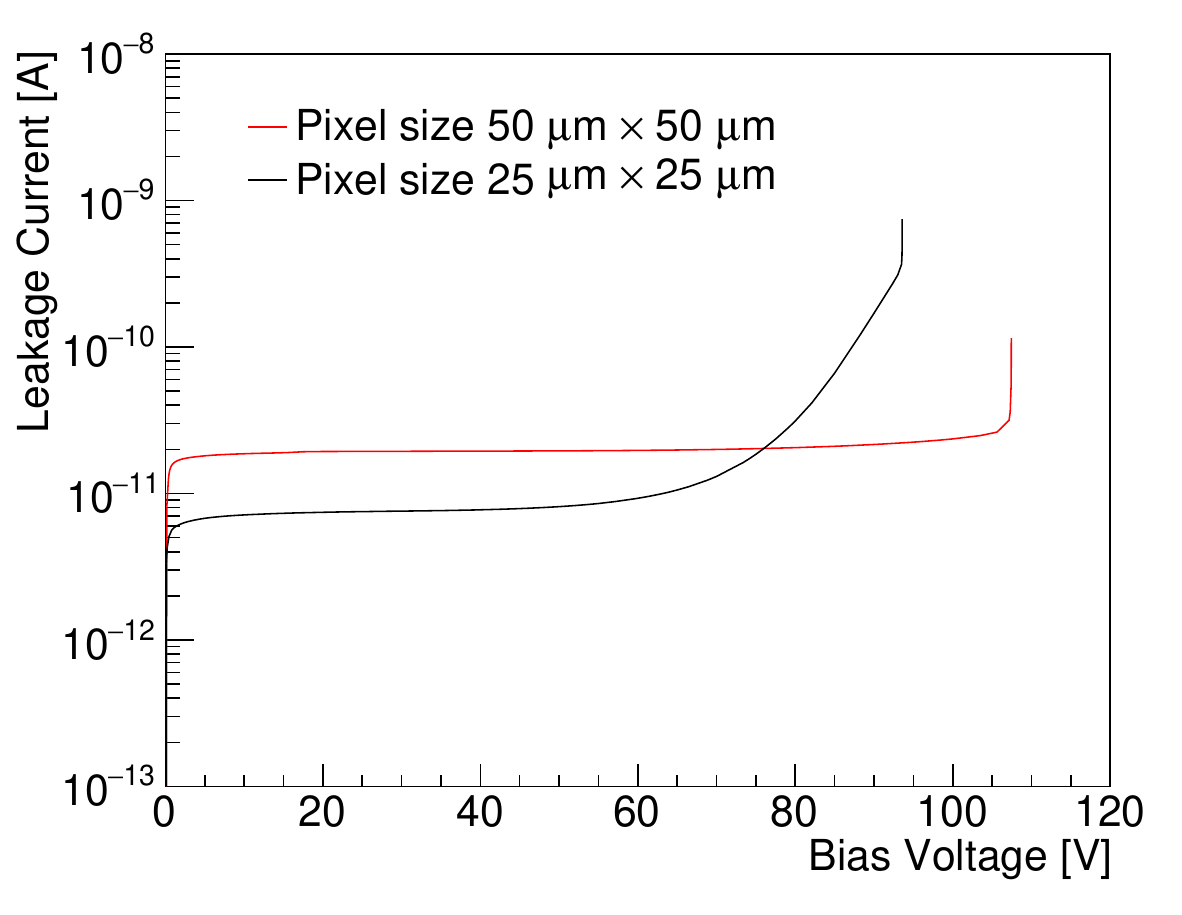}\label{SimulatedIV}}
    \subfigure[]{\includegraphics[width=2.6in]{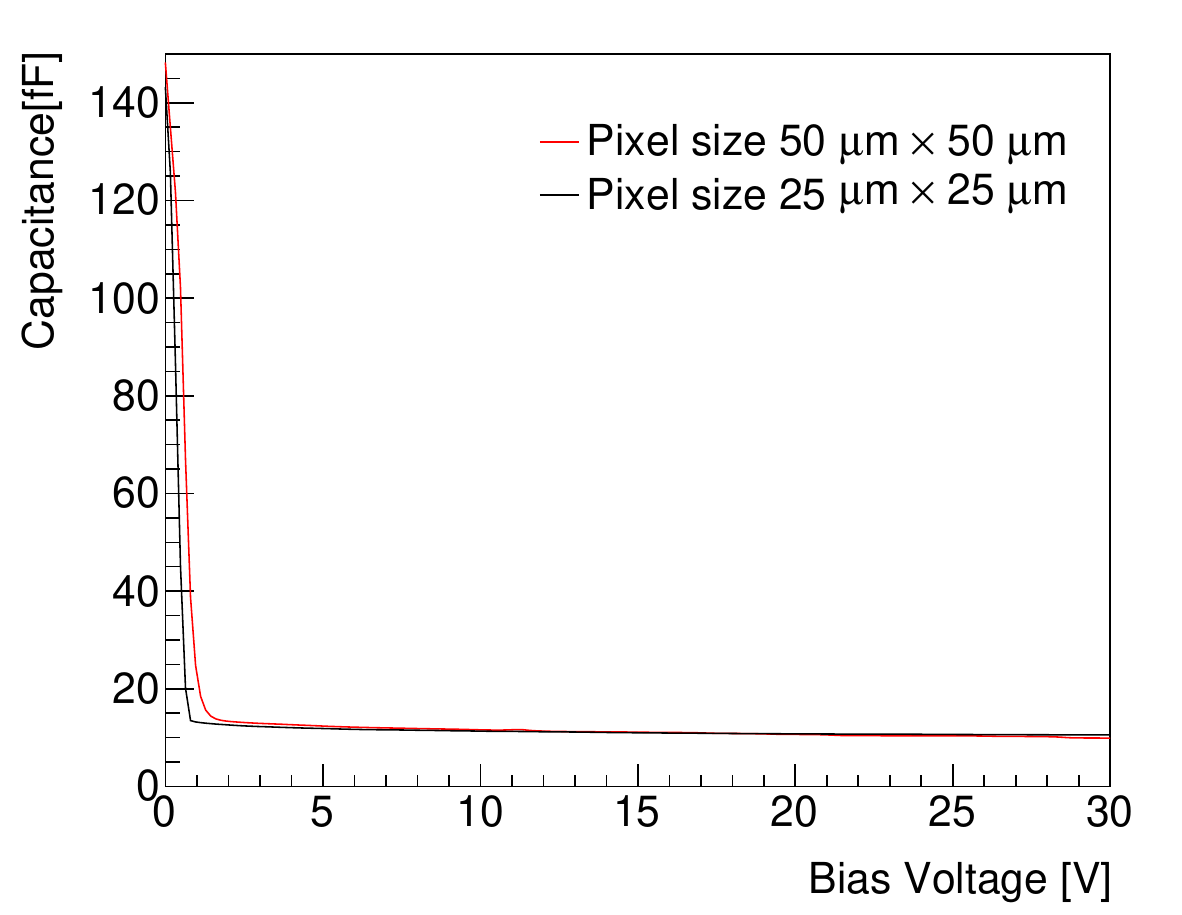}\label{SimulatedCV}}
    \caption{Simulated (a) leakage current versus bias voltage curves and (b) capacitance versus bias voltage curves. All simulations were conducted for the detector operating at a temperature of 20~$^\circ\mathrm{C}$ and the frequency used to simulate the capacitance was 10 kHz.}
    \label{simulatedivcv}
\end{figure}

Fig.~\ref{simulatedivcv} shows the simulated results of the structures with pixel sizes of 50 \SI{}{\micro m}~$\times$~50 \SI{}{\micro m} and 25 \SI{}{\micro m}~$\times$~25 \SI{}{\micro m}. In Fig.~\ref{SimulatedIV}, the $I-V$ curves show the lower saturated leakage current and lower breakdown voltage for the structure with 25 \SI{}{\micro m}~$\times$~25 \SI{}{\micro m} pixel size. Fig.~\ref{SimulatedCV} shows that the lateral depletion voltage of 50 \SI{}{\micro m}~$\times$~50 \SI{}{\micro m} pixel size is approximately 2 V, which is a little larger than that of 25 \SI{}{\micro m}~$\times$~25 \SI{}{\micro m} pixel size (about 1 V), as expected due to the geometries. 

\begin{figure}[htbpt]
    \centering
    \subfigure[]{\includegraphics[width=3.2in]{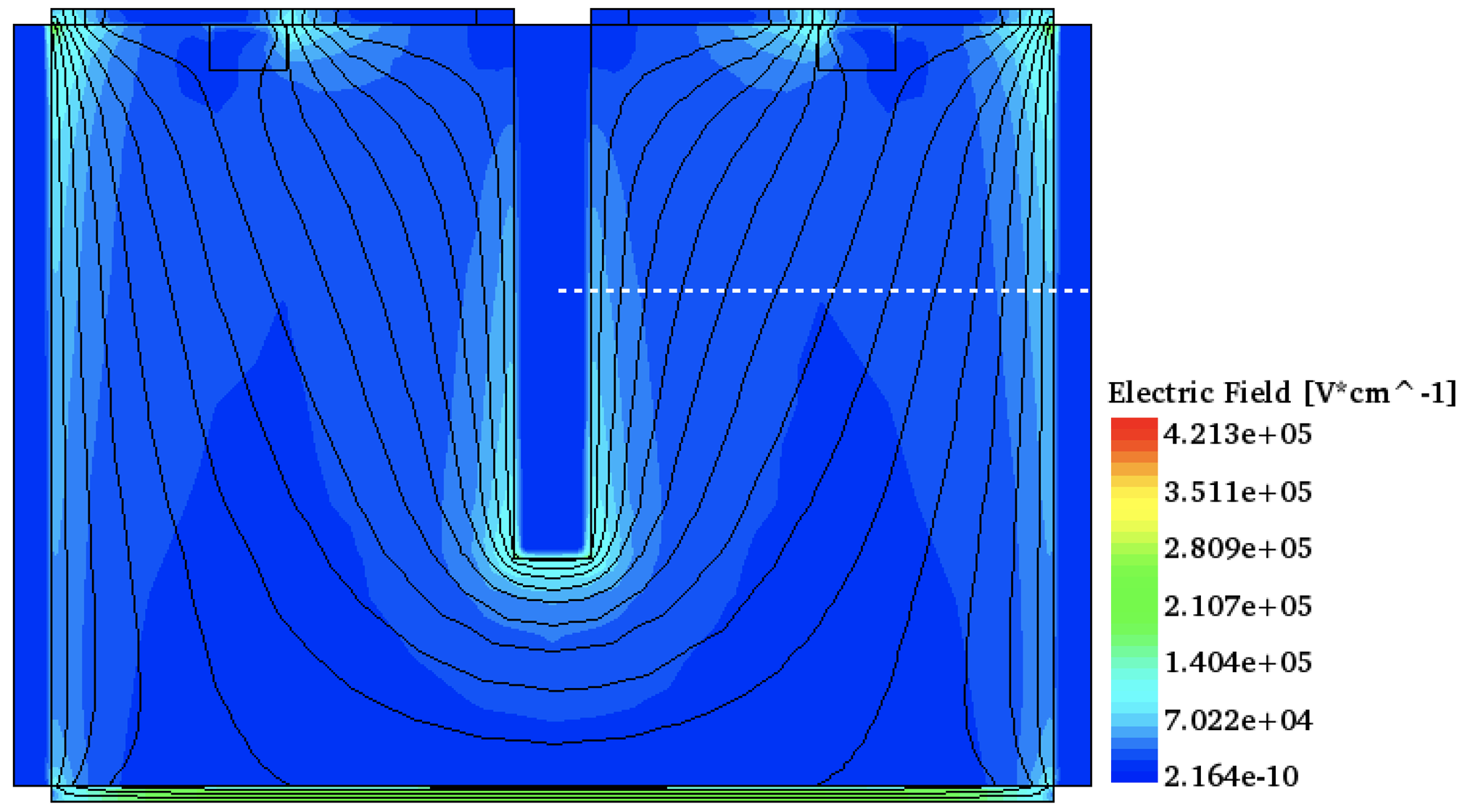}\label{E-Field2d_pitch50_70v}}
    \subfigure[]{\includegraphics[width=2.1in]{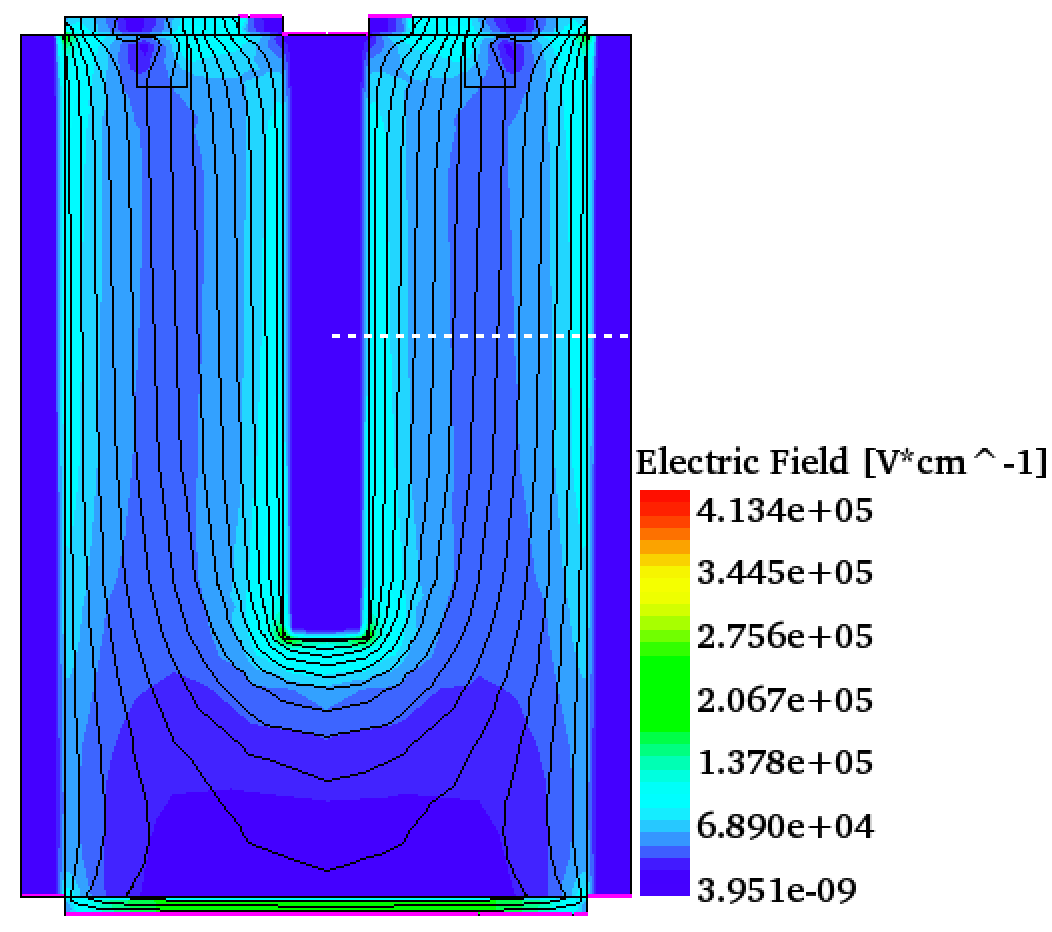}\label{E-Field2d_pitch25_70v}}
    \caption{Cross sections of the electric field distributions for the 3D sensors with different pixel sizes: (a) 50 \SI{}{\micro m} $\times$ 50 \SI{}{\micro m}. (b) 25 \SI{}{\micro m} $\times$ 25 \SI{}{\micro m}. The black lines are equipotential lines whereas the white line is the cutting line. The bias voltage is 70~V.}
    \label{E-Field2d_z37p5}
\end{figure}

\begin{figure}[htbpt]
    \centering
    \subfigure[]{\includegraphics[width=2.6in]{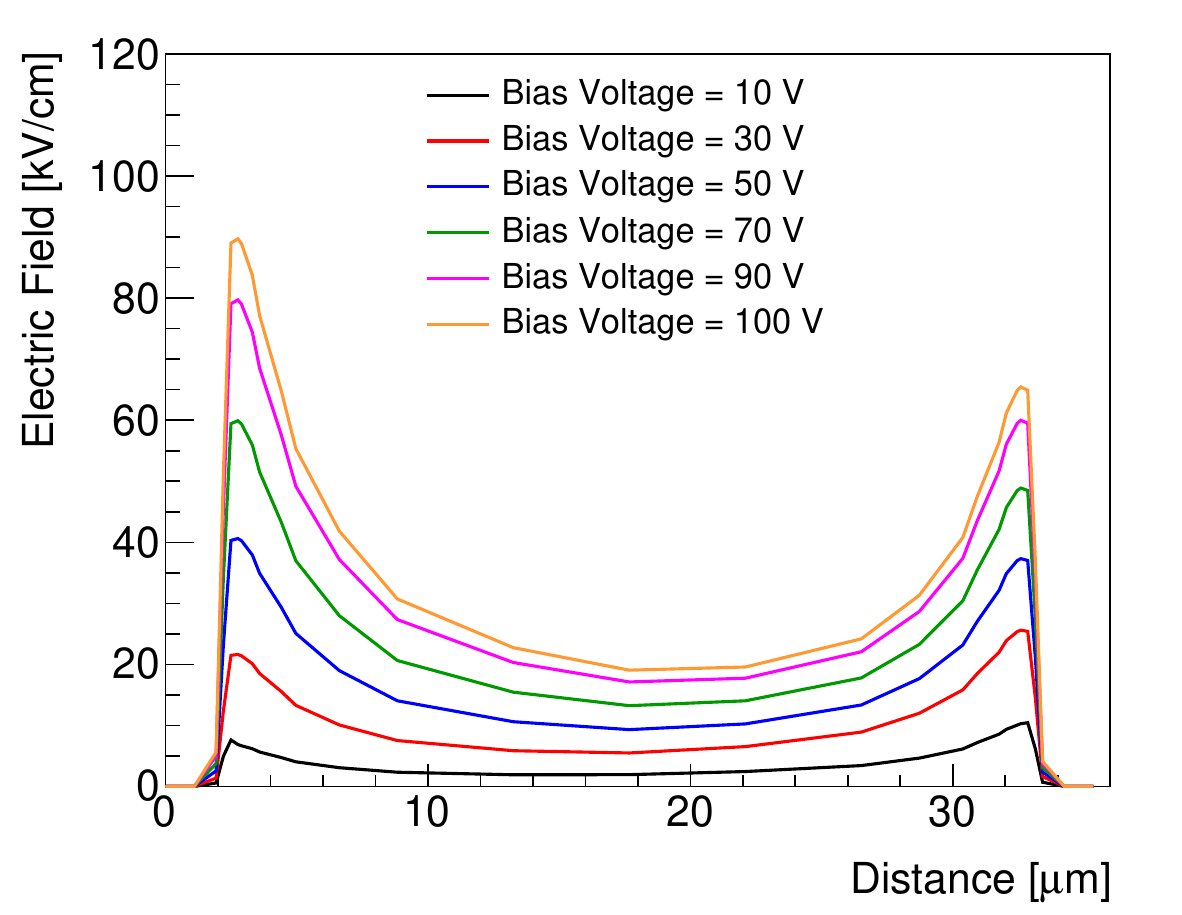}\label{E-Field_pitch50_z37p5}}
    \subfigure[]{\includegraphics[width=2.6in]{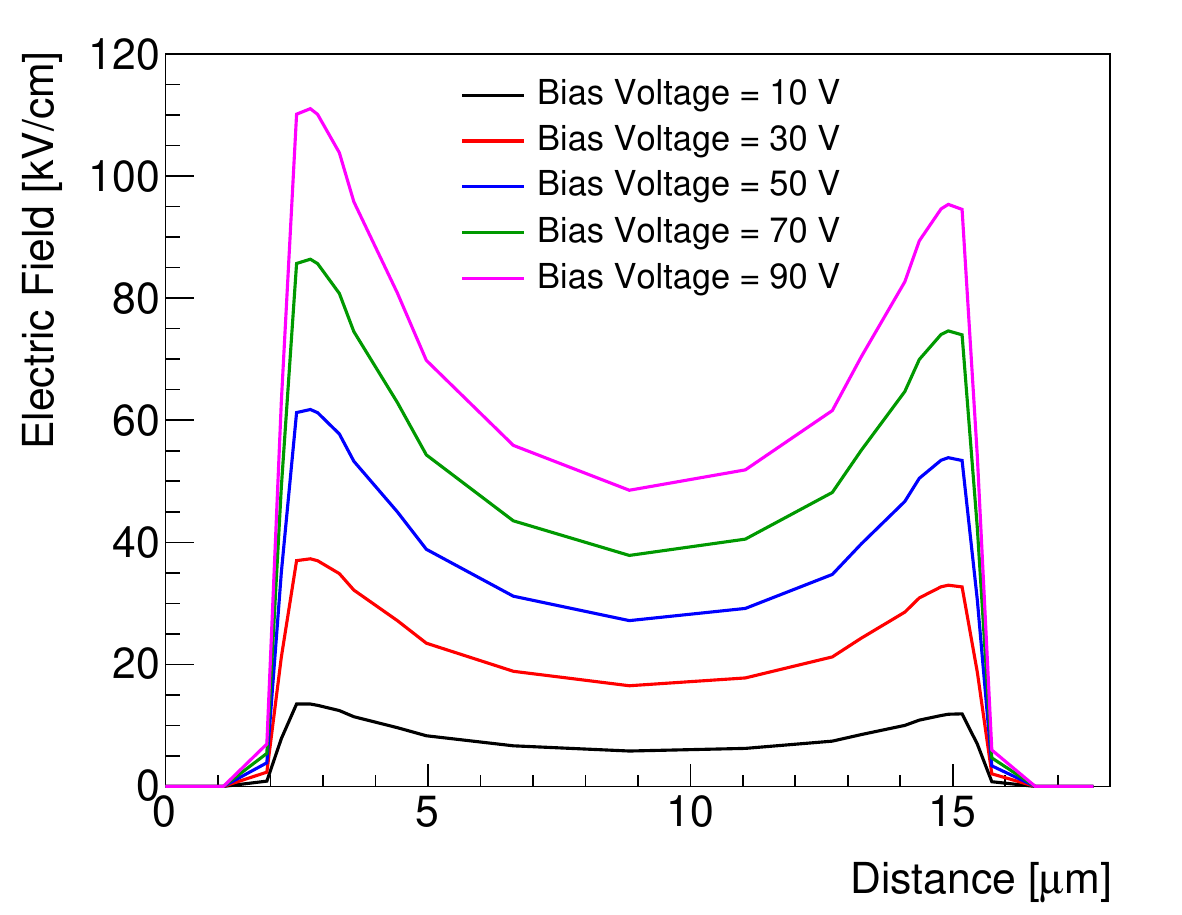}\label{E-Field_pitch25_z37p5}}
    \caption{The electric field distributions between p$^+$ column and n$^+$ column along the white line (see Fig.~\ref{E-Field2d_z37p5}): (a) pixel size of 50 \SI{}{\micro m} $\times$ 50 \SI{}{\micro m}. (b) pixel size of 25 \SI{}{\micro m} $\times$ 25 \SI{}{\micro m}.}
    \label{E-Field_z37p5}
\end{figure}

The electric field distributions between p$^+$ column and n$^+$ column are depicted in Fig.~\ref{E-Field2d_z37p5} when the bias voltage is 70~V. Due to the non-passing-through n$^{+}$ column, the sensors have a low electric field ($<$ 20~kV/cm) region at the bottom. To clarify the behavior of the electric field between p$^+$ column and n$^+$ column at different bias voltages, the electric field distributions along the white line (see Fig.~\ref{E-Field2d_z37p5}) are shown in Fig.~\ref{E-Field_z37p5}. The low electric field region between p$^+$ column and n$^+$ column still exists for the sensor with the pitch of 50 \SI{}{\micro m} at 100~V. It disappears for the sensors with the pitch of 25 \SI{}{\micro m} when the bias voltage reaches 50~V. 

\section{Layout design and fabrication}
\label{Fabrication}

\subsection{Layout design}

\begin{figure}[htbpt]
    \centering
    \subfigure[]{\includegraphics[width=2.6in]{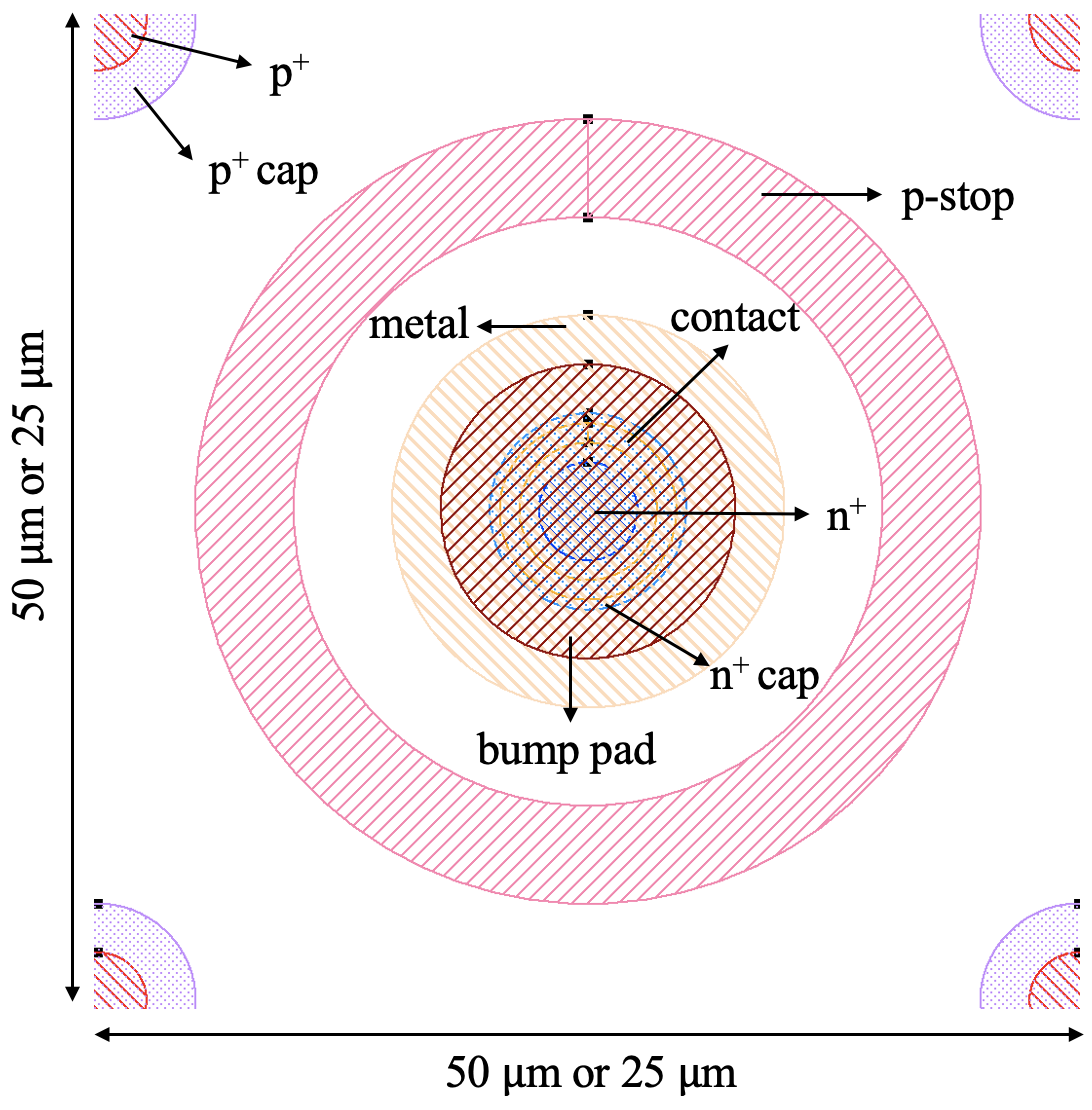}\label{Layout50-7s}}
    \subfigure[]{\includegraphics[width=2.6in]{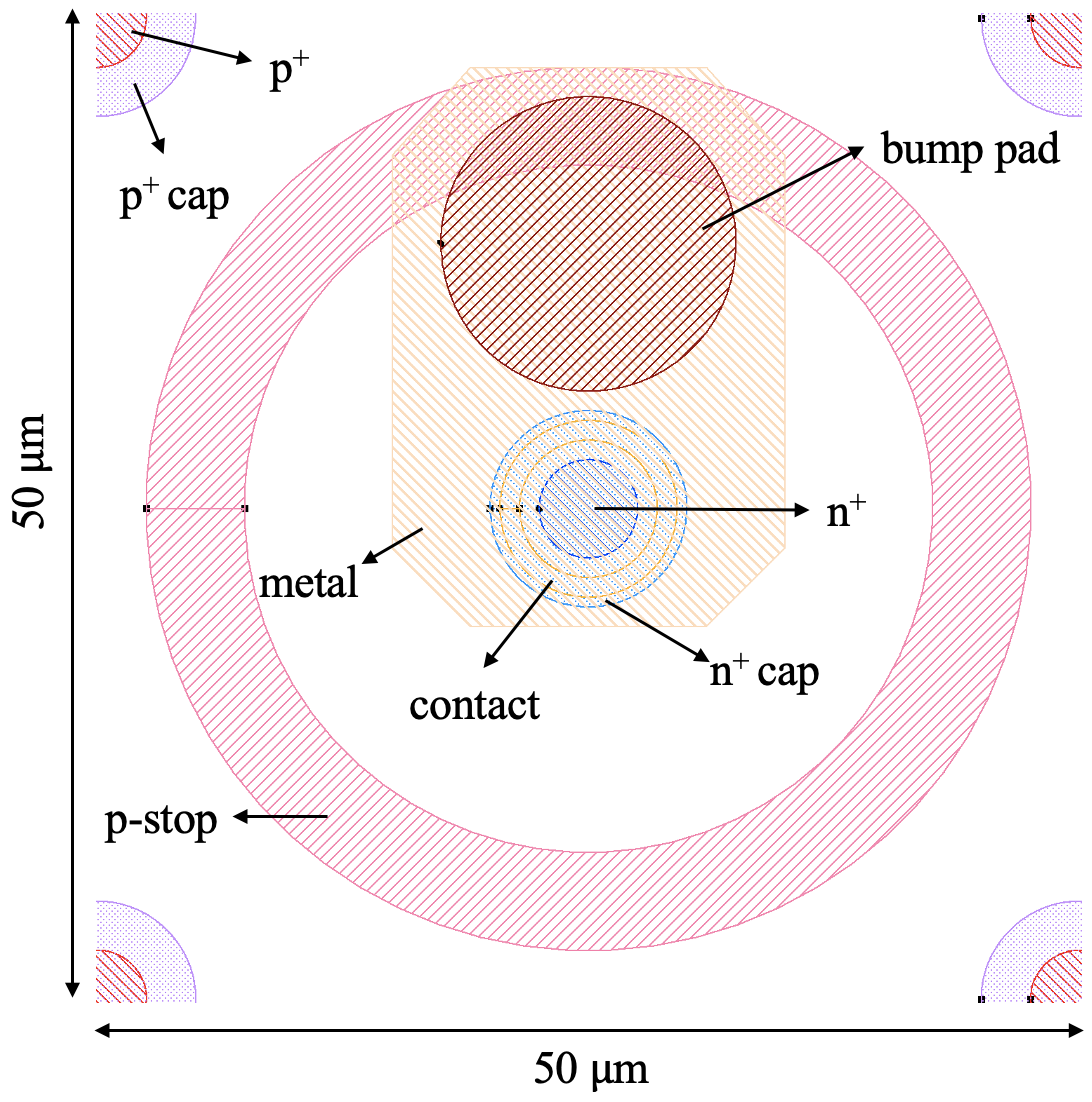}\label{Layout50-7b}}
    \caption{Layout of 3D sensor with 50 \SI{}{\micro m} or 25 \SI{}{\micro m} pitch: (a) bump pad in the center and (b) bump pad separated from the n$^+$ electrode.}
    \label{layout_single}
\end{figure}

In order to fabricate the 3D sensors, a merged layout with different geometries and nine photolithography layers has been designed. 3D sensors with two pitches have a similar basic cell layout, as shown in Fig.~\ref{Layout50-7s}, where the bump pad is positioned in the center. Fig.\ref{Layout50-7b} shows another layout of the 3D sensor with 50 \SI{}{\micro m} pitch, allowing the bump pad to be separated from the n$^+$ electrode. The advantage of the latter design is to place the bump pad in flat areas away from the contact region. There is an oxide layer covering the p-stop ring to avoid shortage between it and the bump pad.

\begin{figure}[htbpt]
    \centering
    \subfigure[]{\includegraphics[width=2.6in]{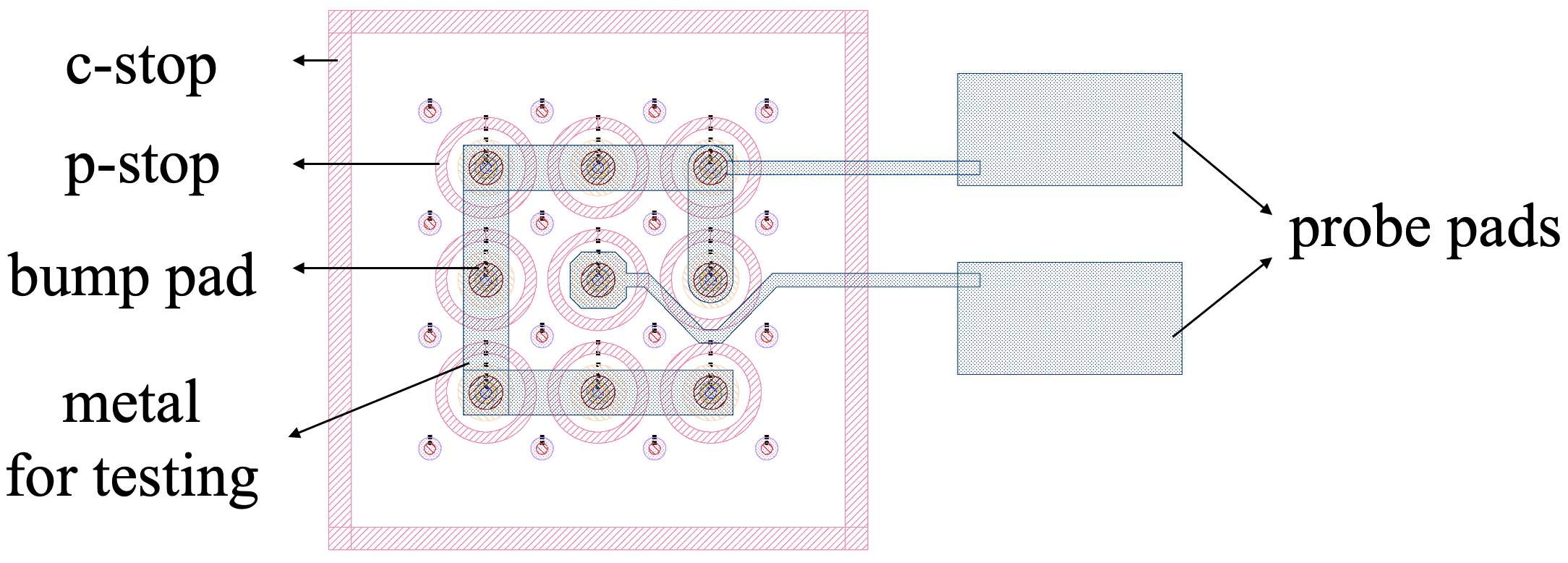}\label{Layout50-7s3c}}
    \subfigure[]{\includegraphics[width=2.6in]{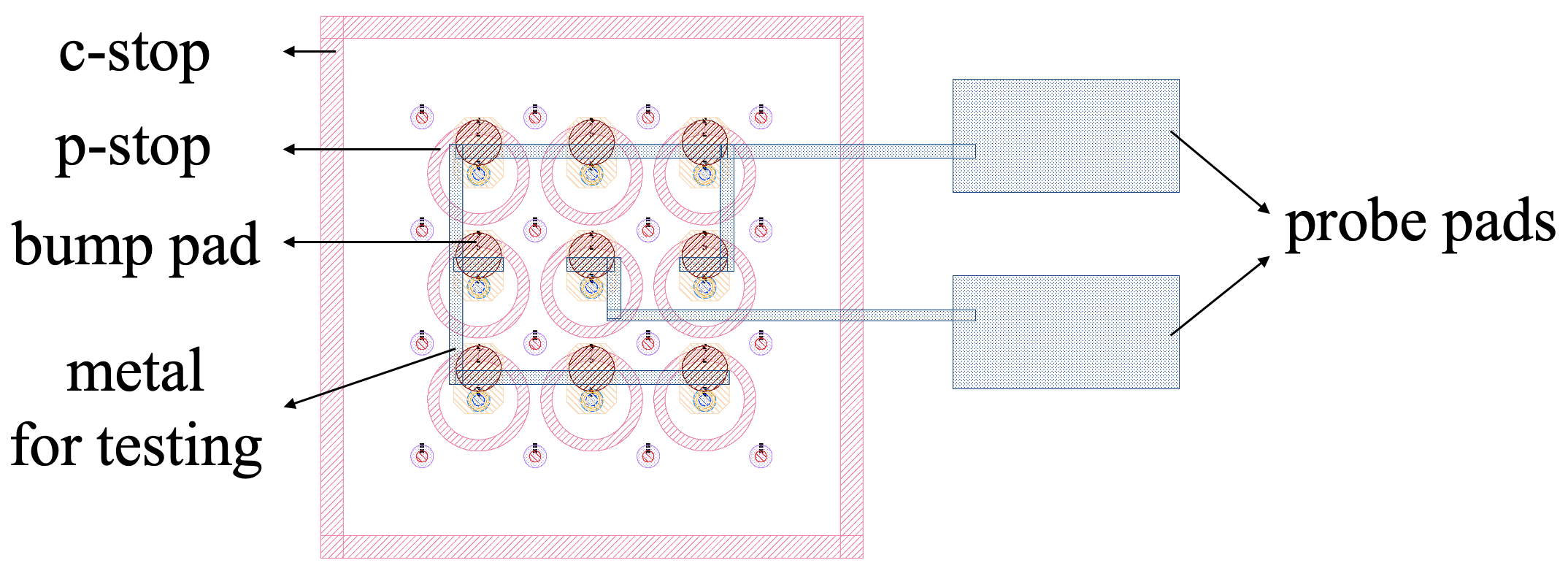}\label{Layout50-7b3c}}
    \subfigure[]{\includegraphics[width=2.6in]{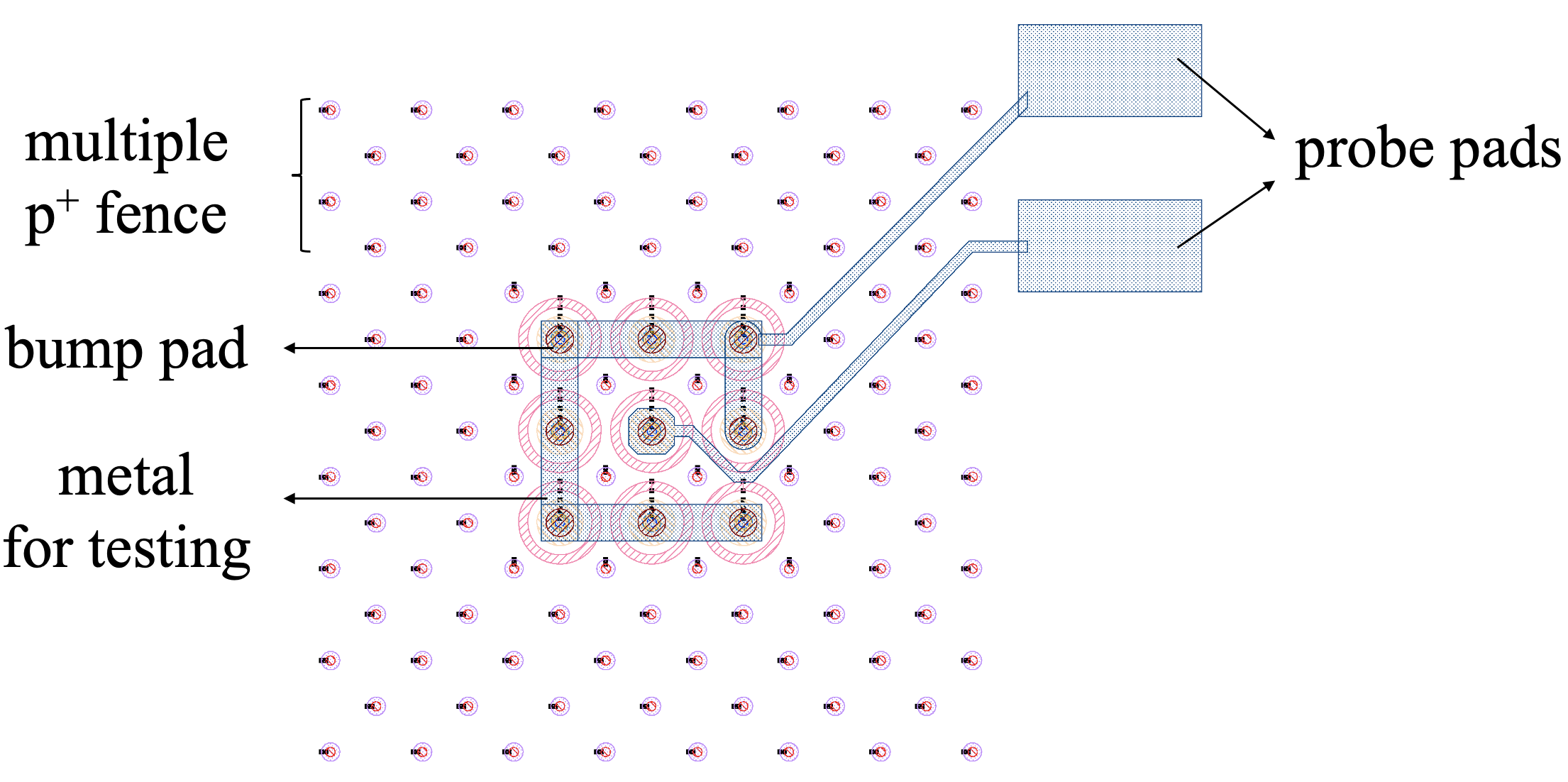}\label{Layout50-7s3m}}
    \subfigure[]{\includegraphics[width=2.6in]{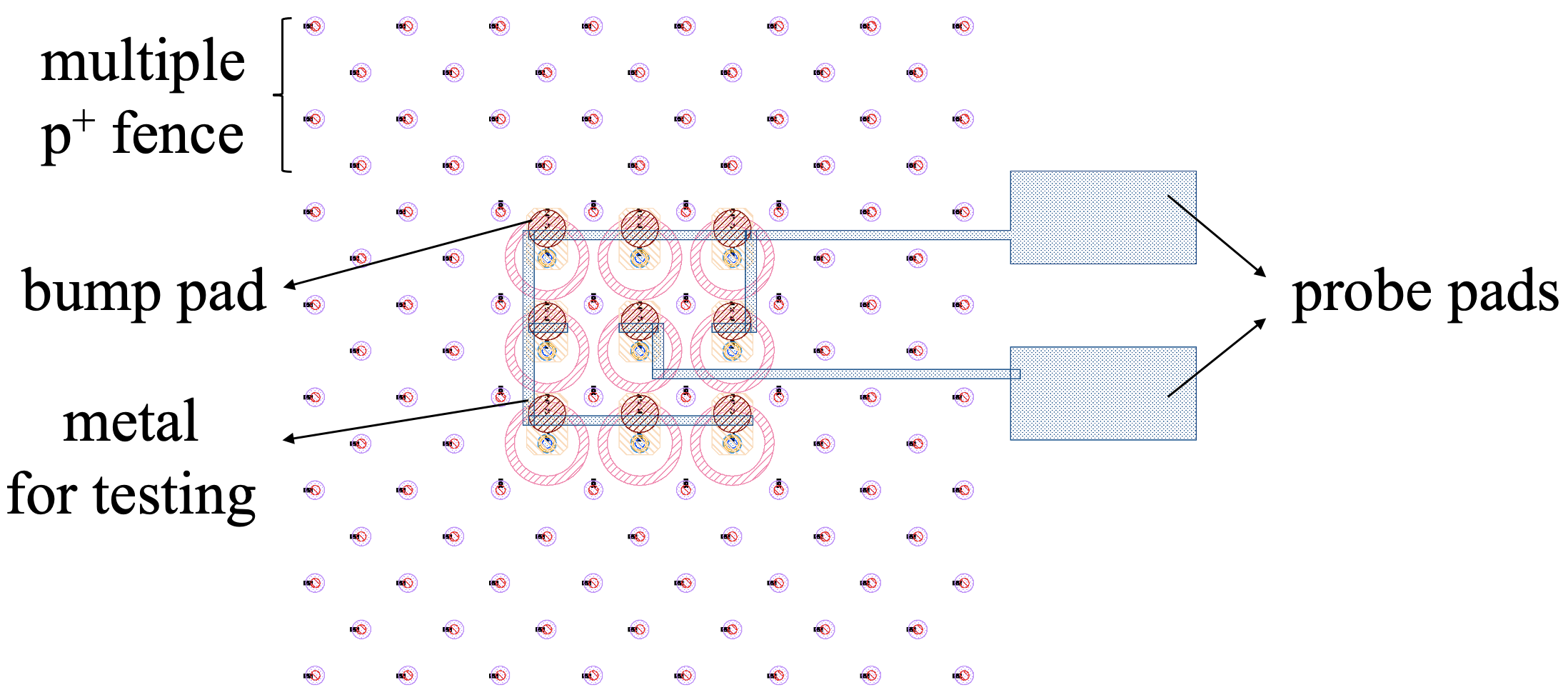}\label{Layout50-7b3m}}
    \caption{Layout of 3~$\times$~3 array sensor with two different edge designs: pixels surrounded by a c-stop (a) bump pad in the center and (b) bump pad separated from the n$^+$ electrode; edge consists of a multiple p$^+$ fence (c) bump pad in the center and (d) bump pad separated from the n$^+$ electrode.}
    \label{layout50_3x3}
\end{figure}

\begin{figure}[htbpt]
    \centering
    \subfigure[]{\includegraphics[width=2.9in]{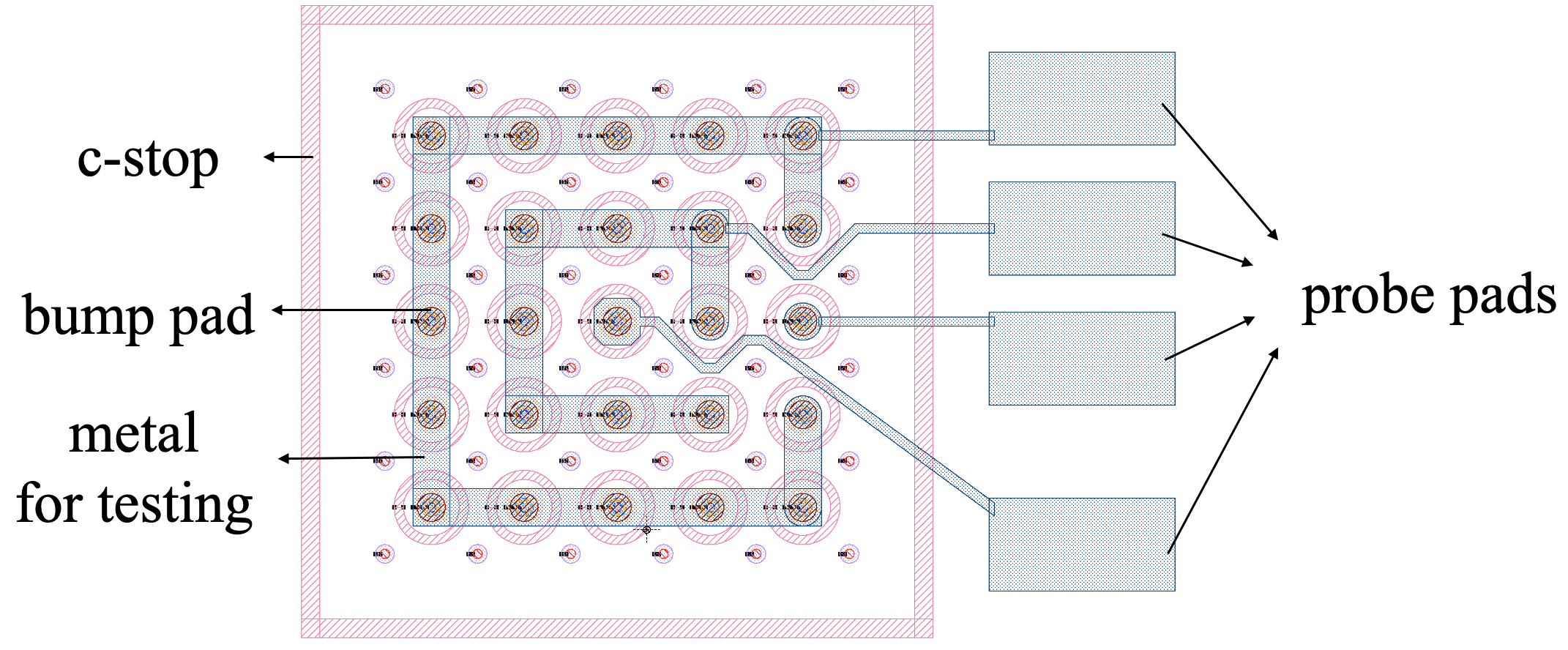}\label{Layout50-7s5c}}
    \subfigure[]{\raisebox{-0.1cm}{\includegraphics[width=2.2in]{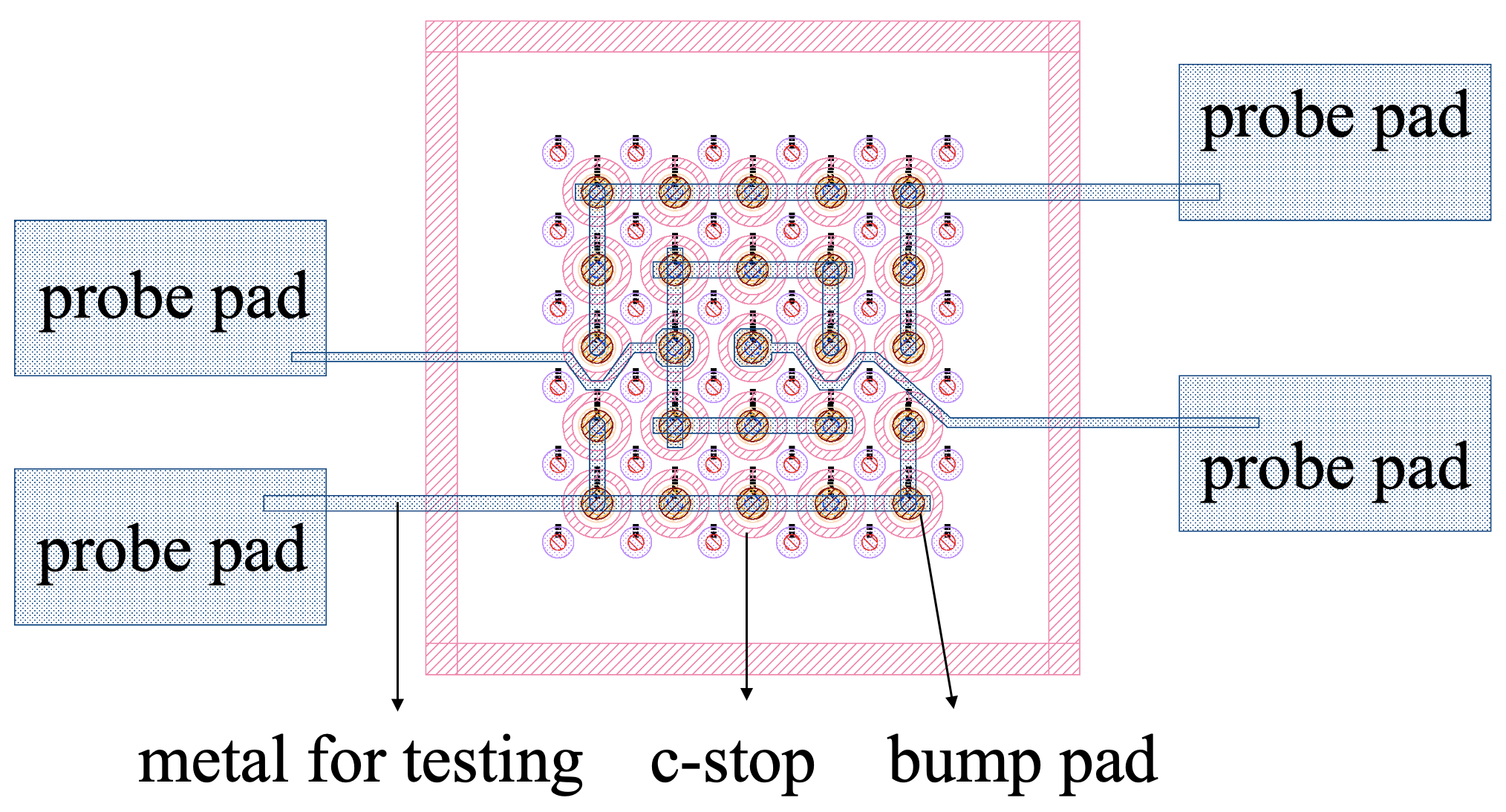}}\label{Layout25-7s5c}}
    \caption{Layout of 5~$\times$~5 array sensor: (a) 50 \SI{}{\micro m} pitch and (b) 25 \SI{}{\micro m} pitch.}
    \label{layout_5x5}
\end{figure}

\begin{figure}[htbpt]
    \centering
    \subfigure[]{\includegraphics[width=2.6in]{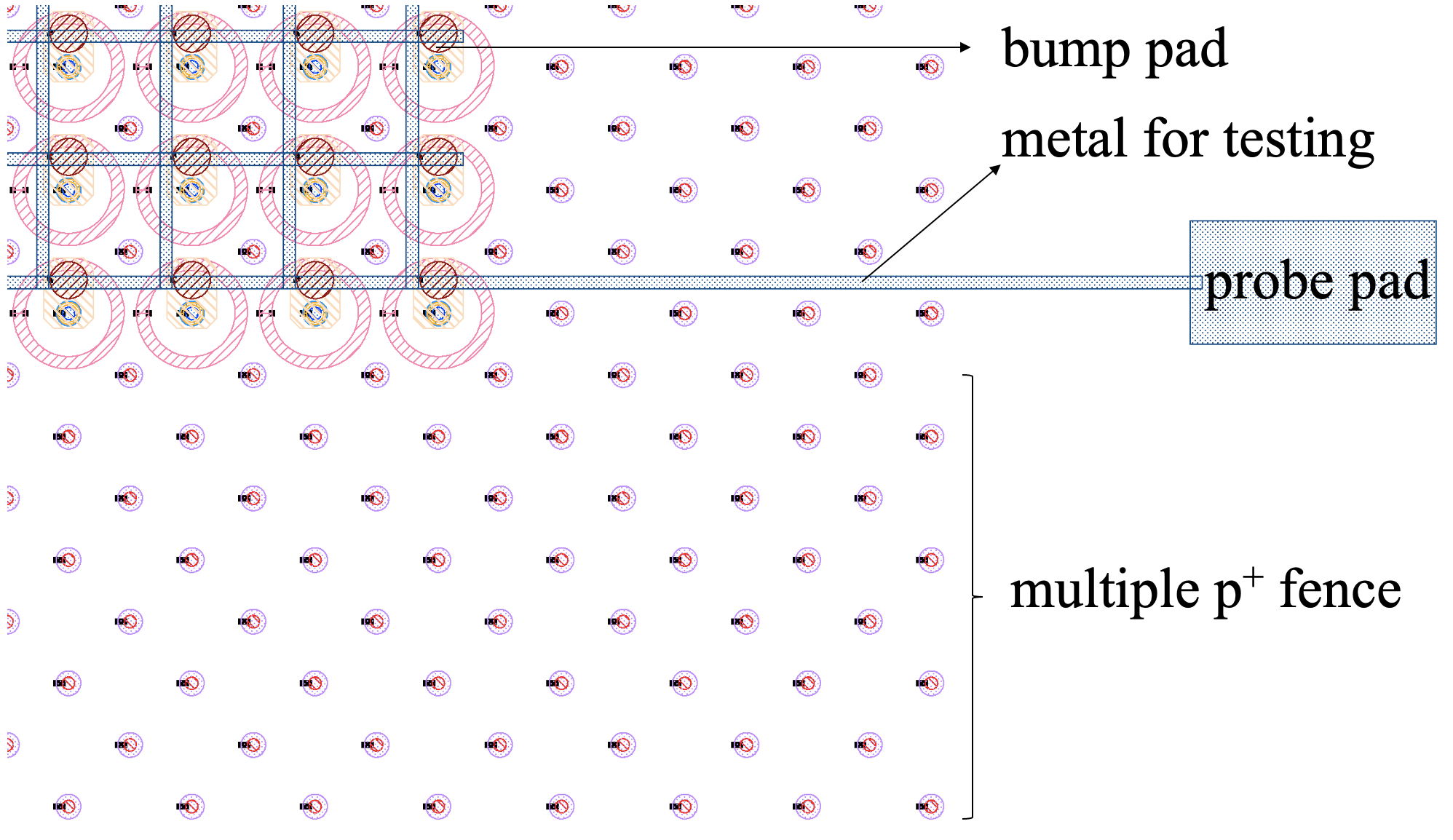}\label{Layout50-l40mpf}}
    \subfigure[]{\includegraphics[width=2.6in]{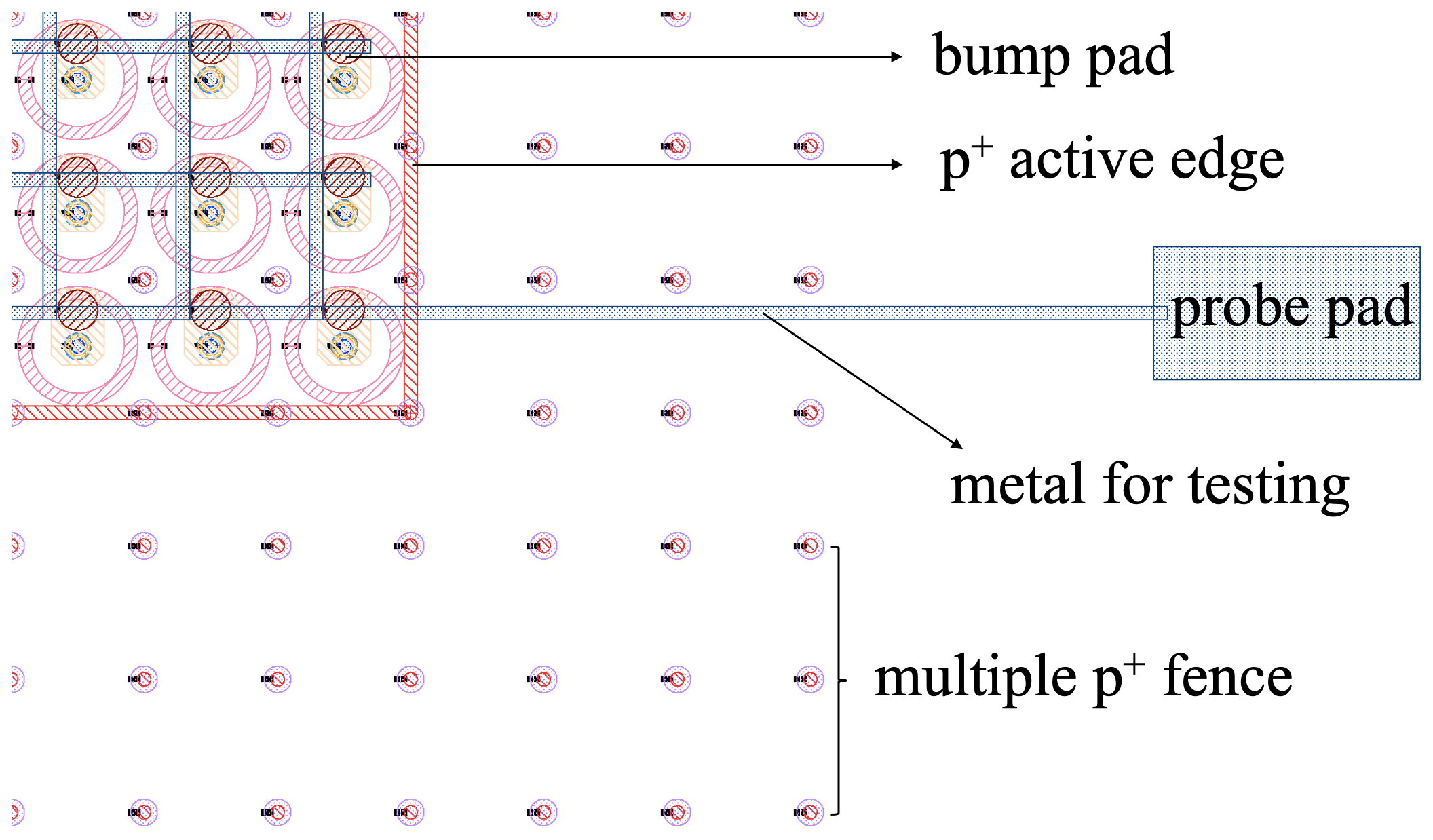}\label{Layout50-l40pa}}
    \caption{Layout of edge designs for 40~$\times$~40 array sensor: edge consists of a multiple p$^+$ fence (a) and a p$^+$ active edge (b).}
    \label{layout_40x40}
\end{figure}

Based on the basic cell, test structures composed of different sizes are designed in the first batch, including 3~$\times$~3, 5~$\times$~5 and 40~$\times$~40 arrays. Due to the small size of the electrodes, an extra metal layer is added to easily characterize these structures. As for the edge, three designs are introduced. For 3~$\times$~3 array sensors, two edge designs are shown in Fig.~\ref{layout50_3x3}. One design is to surround the pixels with a c-stop, that can be implemented with similar process as the p-stop. The c-stop structure is to isolate the active area at the surface. Another design is to achieve a slim edge that consists of a multiple p$^+$ column fence, as described in~\cite{Povoli:2012zz,Sultan:2016vzg}. The 5~$\times$~5 array sensor only uses the c-stop edge design to simplify the layout (see Fig.~\ref{layout_5x5}). The 40~$\times$~40 array sensor is designed to test the capacitance and estimate the yield of devices with a total area of the order of mm$^2$ (approximately 4 mm$^2$ for 50 \SI{}{\micro m} pitch and approximately 1 mm$^2$ for 25 \SI{}{\micro m} pitch). In addition to the slim edge with a multiple p$^+$ fence, a p$^+$ active edge is designed, which is achieved by replacing the outmost ring of p$^+$ columns with a p$^+$ trench, shown in Fig.~\ref{layout_40x40}. Because the quality of the trench is not very clear, the slim edge fence is also added around the trench for protection. The layout designs are summarized in Table~\ref{table:3dlayout.}.


\begin{table}[htbp]
\centering
\caption{Summary of the layout designs.}
\label{table:3dlayout.}
\renewcommand{\arraystretch}{1.2}
\begin{tabular}{cccc}
    \toprule
    Pitch size & Array & Edge design & Bump pad position \\
    \midrule
    \multirow{8}{*}{50 \SI{}{\micro m}} & \multirow{4}{*}{3 $\times$ 3} & \multirow{2}{*}{c-stop} & on the n$^+$\\
     &  &  & separated from the n$^+$\\
    \cline{3-4}
     &  & \multirow{2}{*}{multiple p$^+$ column fence} & on the n$^+$\\
     &  &  & separated from the n$^+$\\
    \cline{2-4}
     & \multirow{2}{*}{5 $\times$ 5} & \multirow{2}{*}{c-stop} & on the n$^+$\\
     &  &  & separated from the n$^+$\\
    \cline{2-4}
     & \multirow{2}{*}{40 $\times$ 40} & multiple p$^+$ column fence & \multirow{2}{*}{separated from the n$^+$}\\
     &  & p$^+$ active edge & \\
    \hline
    \multirow{5}{*}{25 \SI{}{\micro m}} & \multirow{2}{*}{3 $\times$ 3} & c-stop & \multirow{5}{*}{on the n$^+$}\\
     &  & multiple p$^+$ column fence & \\
     & 5 $\times$ 5 & c-stop & \\
     & \multirow{2}{*}{40 $\times$ 40} & multiple p$^+$ column fence & \\
     &  & p$^+$ active edge & \\
    \bottomrule
\end{tabular}
\end{table}

\subsection{Fabrication}
\label{fabrication}

The overall process of small-pitch 3D sensors have been reproted in~\cite{DallaBetta:2016czx}. However, the detailed process needs to be altered for different fabrication facilities. This section details the fabrication process for the 3D sensors, developed and optimized in the clean room of the USTC Center for Micro and Nanoscale Research and Fabrication (USTC-NRFC).

\begin{figure}[htbp]
    \centering
    \subfigure[]{\includegraphics[width=2.3in]{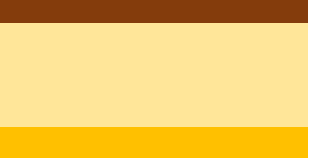}\label{FO}}
    \hspace{1cm}
    \subfigure[]{\includegraphics[width=2.3in]{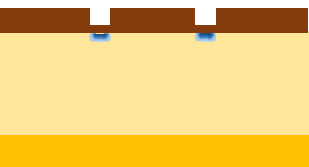}\label{PS}}
    \subfigure[]{\includegraphics[width=2.3in]{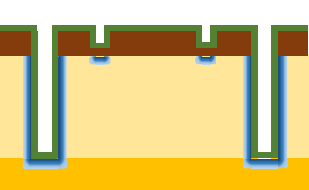}\label{PCol}}
    \hspace{1cm}
    \subfigure[]{\includegraphics[width=2.3in]{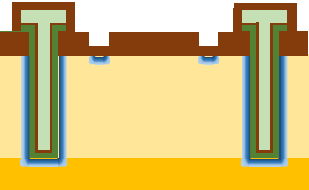}\label{PCap}}
    \subfigure[]{\includegraphics[width=2.3in]{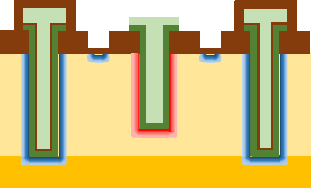}\label{NCap}}
    \hspace{1cm}
    \subfigure[]{\includegraphics[width=2.3in]{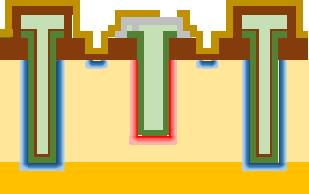}\label{Pass}}
    \subfigure[]{\includegraphics[width=2.3in]{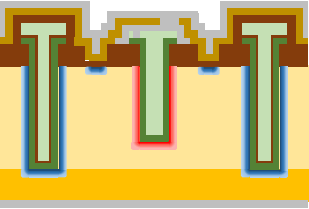}\label{Final}}
    \hspace{1cm}
    \subfigure{\includegraphics[width=2.3in]{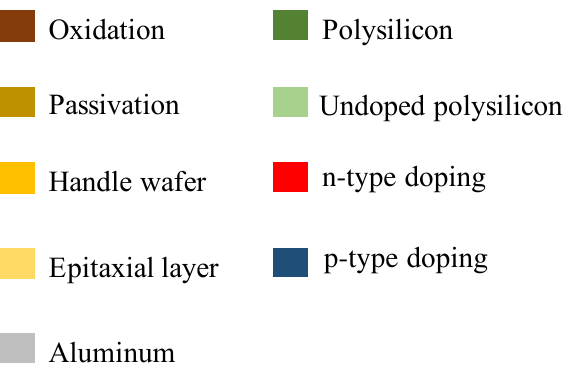}\label{LABEL}}    
    \caption{Main steps for the sensor on p-type substrate (not to scale).}
    \label{Schematicdiagram}
\end{figure}

Sensors were fabricated on 6-inch (150 mm diameter), p-type, (100) oriented epitaxial silicon wafers, with a total thickness of 675 \SI{}{\micro m}. For the epitaxial layer, the thickness is 50 \SI{}{\micro m} and the nominal resistivity is greater than 1000 $\Omega~\cdot$~cm. The fabrication process requires nine photolithography mask plates; and more than 100 steps are carried out. With the help of schematic cross-sectional diagrams in Fig.~\ref{Schematicdiagram}, the main process steps are summarized as follows. 

\begin{enumerate}
    \item[a)] A thermal dioxide is grown by a dry oxidation process (1050~$^\circ\mathrm{C}$ for 600 minutes in an oxygen atmosphere) to be used as the high quality isolation layer. 
    \item[b)] After removal of silicon dioxide on the top of the p-stop/c-stop structure, boron is implanted at an energy of 55 keV and a dose of 4$\times$10$^{14}$ cm$^{-2}$. Following implantation, the wafers are annealed at 1050~$^\circ\mathrm{C}$ for 11 minutes in an atmosphere containing water vapor.
    \item[c)] A 200 nm aluminum film deposited by the electron beam evaporation technique is served as the hard mask for DRIE steps. After the DRIE steps, the p$^+$ column is partially filled by the polysilicon (about 1 \SI{}{\micro m} thick) using the Low Pressure Chemical Vapor Deposition (LPCVD) technique. Boron is diffused at 950~$^\circ\mathrm{C}$ for 40 minutes from a 6-inch boron nitride (BN) wafer in a controlled nitrogen/oxygen atmosphere. 
    \item[d)] A thin oxide layer is thermally grown on the doped polysilicon to prevent dopant out-diffusion and the p$^+$ column is filled by the undoped polysilicon to ensure mechanical robustness. Next, the p$^+$ cap is fabricated by etching the unwanted polysilicon and SiO$_2$ on the surface, and then the p$^+$ column is protected by a thin layer of thermal dioxide.   
    \item[e)] The aluminum is deposited and patterned on the front side again for the n$^+$ column etching. Following the DRIE steps with fewer cycles, the shallower n$^+$ column is partially filled by the polysilicon (about 1 \SI{}{\micro m} thick), and then doped by a 6-inch phosphorus solid source wafer which is made of quartz and silicon phosphate. The diffusion procedure takes place at 950~$^\circ\mathrm{C}$ for 40 minutes in a controlled nitrogen atmosphere. Unlike the p$^+$ column, the undoped polysilicon is directly deposited to fill the n$^+$ column without a thin SiO$_2$ layer.
    \item[g)] The undoped polysilicon of the n$^+$ cap is patterned and etched to define the contact regions. Next, a thin layer of aluminum is deposited uniformly on the surface, and the metal pattern is defined after removing the unprotected aluminum. To protect the sensor from external contamination, the surface is passivated with 1 \SI{}{\micro m} SiO$_2$, which is patterned to expose the metal layer. 
    \item[h)] Finally, the metallization is carried out on both sides. The deposited metal on the backside directly connects the p$^+$ electrodes through the handle wafer with a low resistivity in the range of 0.004 $-$ 0.01 $\Omega~\cdot$~cm. After patterning and etching, the remaining metal on the front side, which shortens the pixels, is used to perform the characterization of the sensors. 
\end{enumerate}

\begin{figure}[htbp]
    \centering
    \includegraphics[width=2.6in]{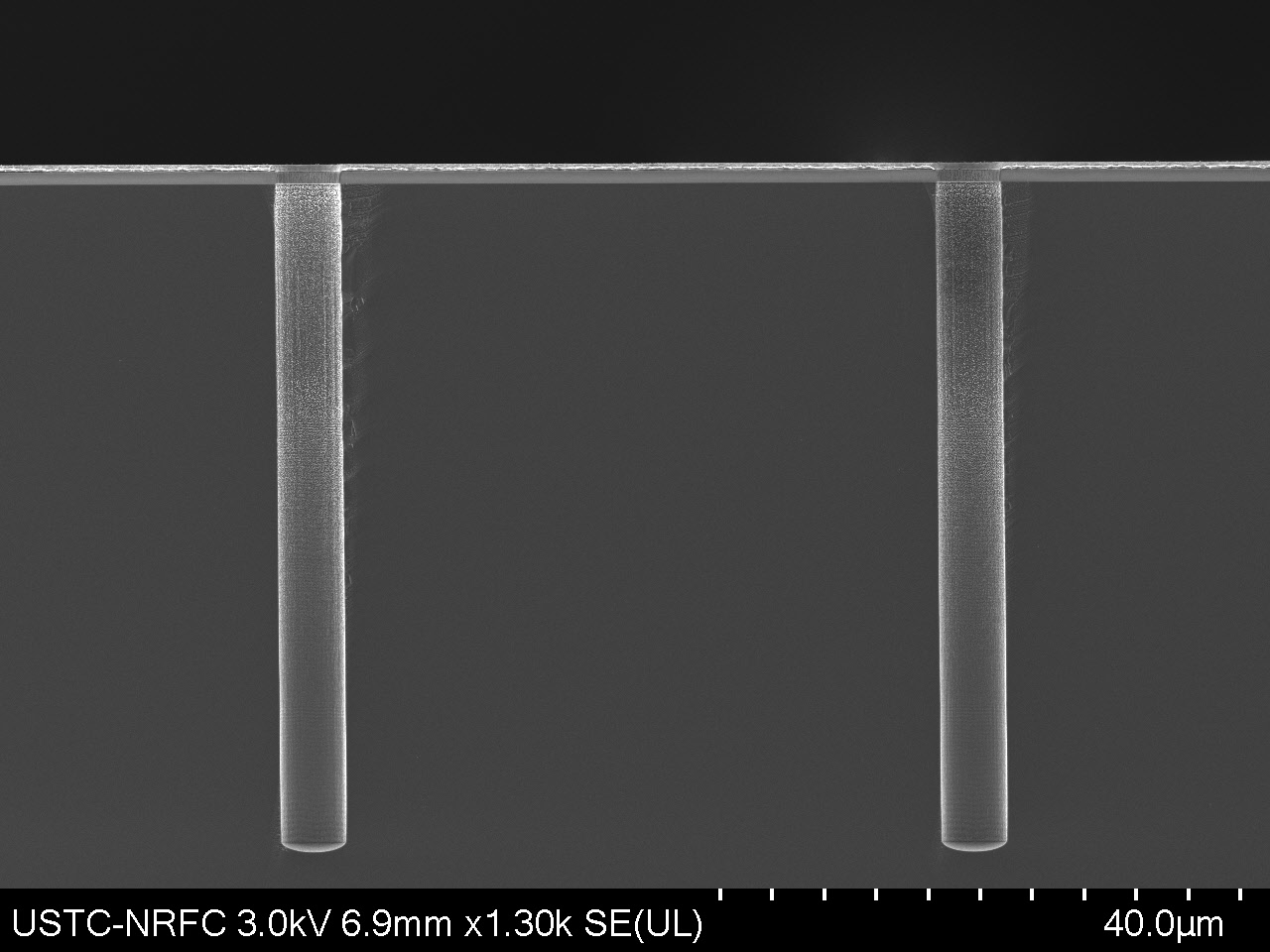}
    \caption{SEM micrograph of columns on a test wafer.}
    \label{DRIE50}
\end{figure}

DRIE is the key technology in the fabrication process. The DRIE technique at NRFC is based on the Bosch process~\cite{bosch} using PlasmaPro 100 Estrelas. While achieving the high aspect ratio (defined as the maximum depth divided by the maximum width) holes, it is crucial to control the vertical profile and sidewall damage. Therefore, a modified procedure of Bosch process has been developed, and the process parameters have been opportunely altered. The standard procedure consists of the repetition of a two-part etching phase, alternating isotropic etching steps and sidewall passivation steps. In the modified procedure, the isotropic etching step is split into two phases: anisotropic etching step for removing the bottom passivation layer and isotropic etching step for etching the silicon substrate. Compared with isotropic etching step, anisotropic etching step uses lower inductive coupled power (ICP) and lower SF$_6$ gas flow to project the sidewall passivation layer as much as possible. Besides, the modified Bosch process is divided into multiple stages and a set of process parameters for every etching stage has been optimized according to the etching depth. Five sets of process parameters were used to etch the p$^{+}$ column, whereas four sets of process parameters and fewer cycles were used to etch the n$^{+}$ column due to the shallower depth. A 4-inch test wafer was etched by the modified Bosch process and the Scanning Electron Microscopy (SEM) micrograph in Fig.~\ref{DRIE50} shows the cross section of the columns, indicating an appropriate process recipe has been established.

Diffusion method is employed to introduce impurity atoms into the holes. In order to develop safe and environmentally friendly diffusion technique, solid source wafers are selected because common non-solid sources, such as gaseous source (B$_2$H$_6$, PH$_3$) and liquid source (BBr$_3$, POCl$_3$), are highly toxic. The solid source wafer is placed in a boat quartz boat between two adjacent silicon wafers and then heated at high temperature in a furnace tube. The formed borosilicate glass (BSG) or phosphosilicate glass (PSG) on silicon wafers operates as the dopant source. After diffusion, the BSG or PSG is removed using the buffered oxide etch (BOE) solution to avoid excessive doping in the subsequent high-temperature process.

\begin{figure}[htbp]
    \centering
    \includegraphics[width=2.8in]{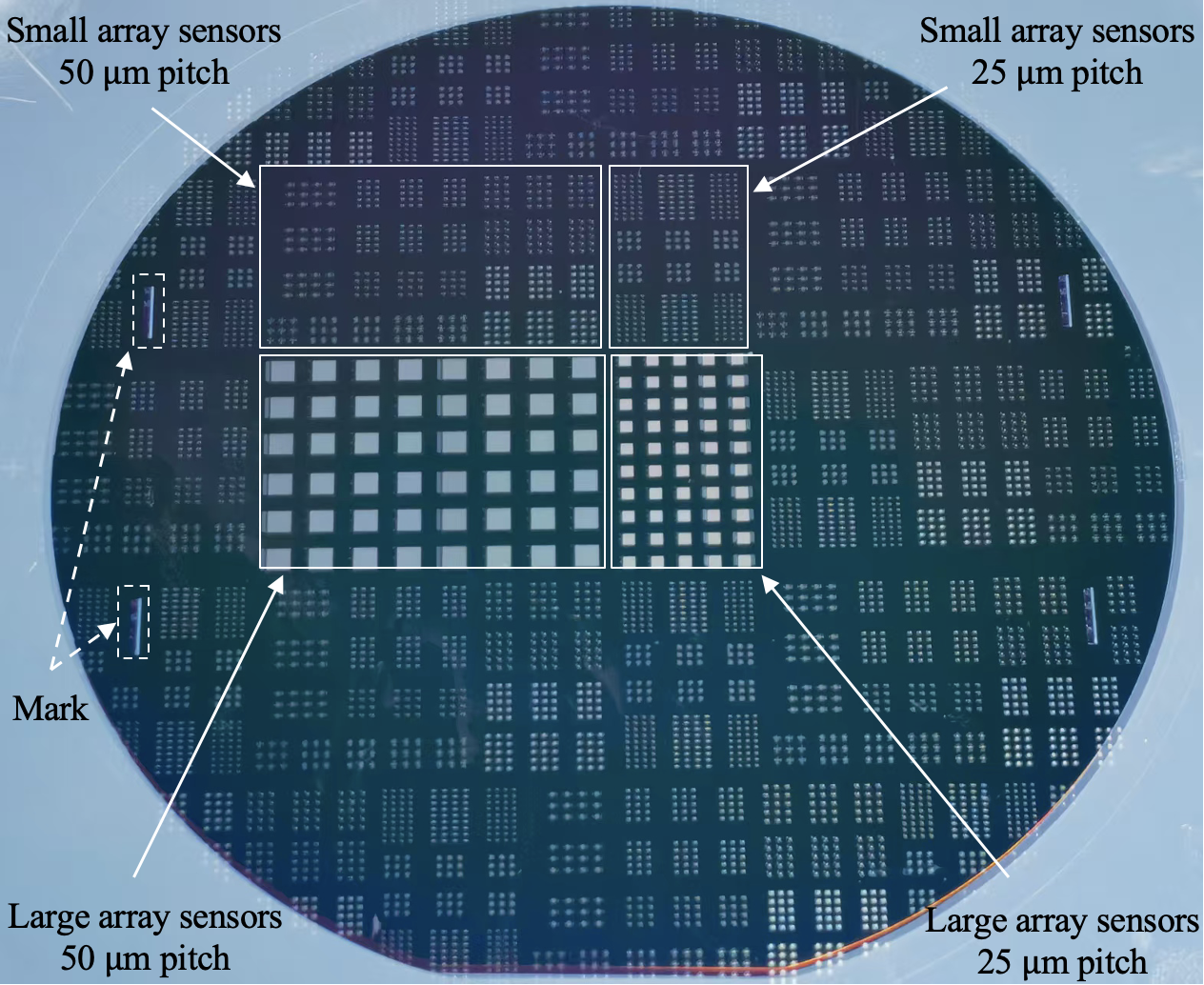}\label{wafer2}
    \caption{Photograph of a fabricated wafer. The large array sensors are placed in the central region, surrounded by the small array sensors. The white boxes enclose all the design types.}
    \label{wafer}
\end{figure}

\begin{figure}[htbp]
    \centering
    \includegraphics[width=2.8in]{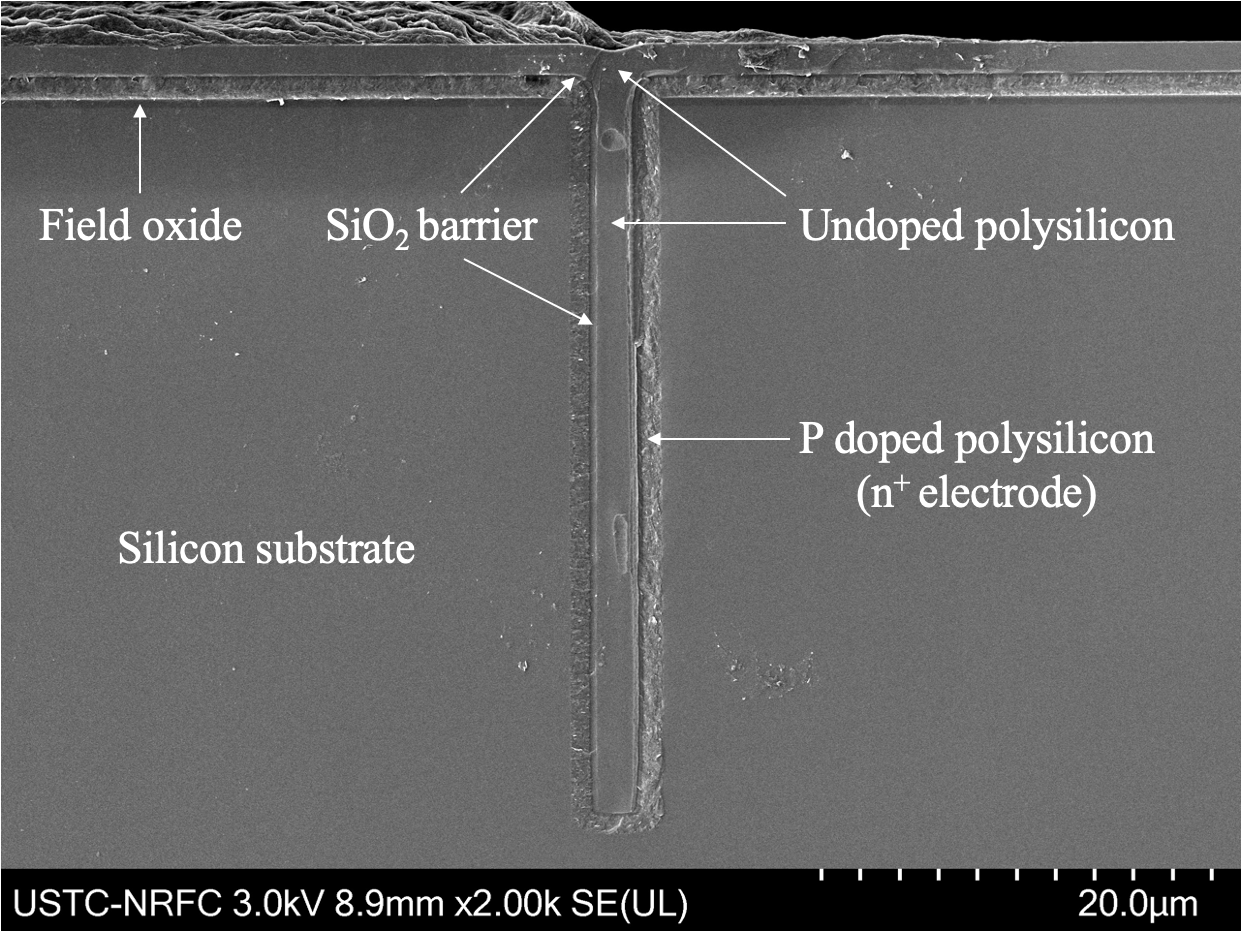}
    \caption{SEM micrograph of a n$^+$ column.}
    \label{filledcolumn}
\end{figure}

Fig.~\ref{wafer} shows the photograph of a fabricated wafer, where the 3D sensors are within 4-inch efficient region. The large array sensors, composed of 40~$\times$~40 pixels, are placed in the central region. Outside the large array sensors, the small array sensors including 3~$\times$~3 and 5~$\times$~5 arrays are arranged compactly to make full use of the wafer. The white boxes enclose all the design types. From the SEM analysis results shown in Fig.~\ref{filledcolumn}, it is confirmed that the complete column has been fabricated successfully.

\section{Measurement Setup}
\label{Measurement Setup}
Two types of measurements have been carried out: leakage current versus bias voltage, and time resolution tested with a $^{90}$Sr beta source.

The $I-V$ measurements setup at USTC comprised a Keithley 2410 High Voltage SourceMeter and a dual-channel pico-ammeter with fA-level resolution (Keithley 6482) coupled to a semi-automatic probe station. During testing, a negative high voltage was applied from the sensor backside and the leakage current was measured through a probe needle that connects the sensor probe pad. For the 3 $\times$ 3 and 5 $\times$ 5 array sensors, the leakage current of the central pixel and its eight connected neighboring pixels were measured simultaneously by two probe needles. All measurements were performed at a temperature of 22$\pm$1~$^{\circ}\mathrm{C}$ and a relative humidity of 38$\pm$5$\%$.

\begin{figure}[htbp]
    \centering
    \subfigure[]{\includegraphics[width=1.7in]{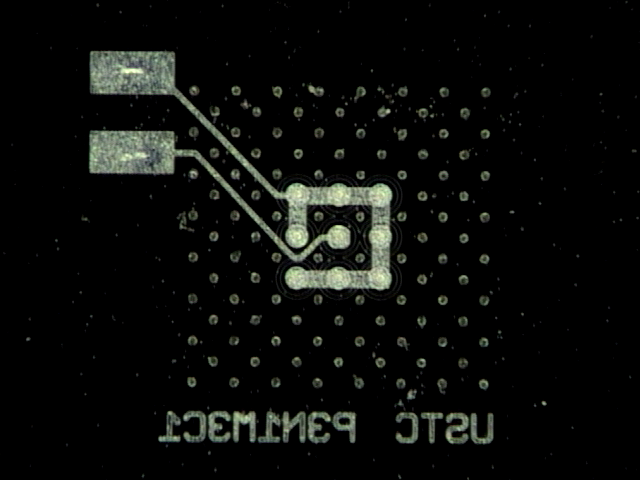}}\label{3D3x3_fence}
    \subfigure[]{\includegraphics[width=1.7in]{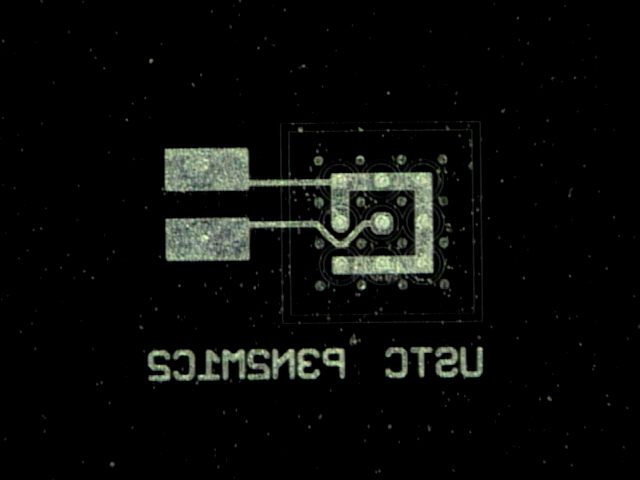}}\label{3D3x3_cstop}
    \subfigure[]{\includegraphics[width=1.7in]{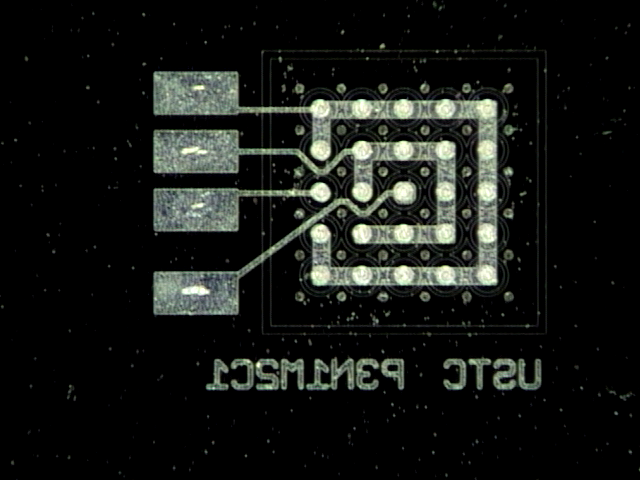}}\label{3D5x5_cstop}
    \caption{Micrograph of the typical 3 $\times$ 3 and 5 $\times$ 5 array sensors with 50 \SI{}{\micro m} pitch: (a) a 3 $\times$ 3 array sensor with a multiple p$^+$ fence, (b) a 3 $\times$ 3 array sensor with a c-stop, and (c) a 5 $\times$ 5 array sensor with a c-stop.}
    \label{ivsetup_3d}
\end{figure}

\begin{figure}[htbp]
    \centering
    \includegraphics[width=5.0in]{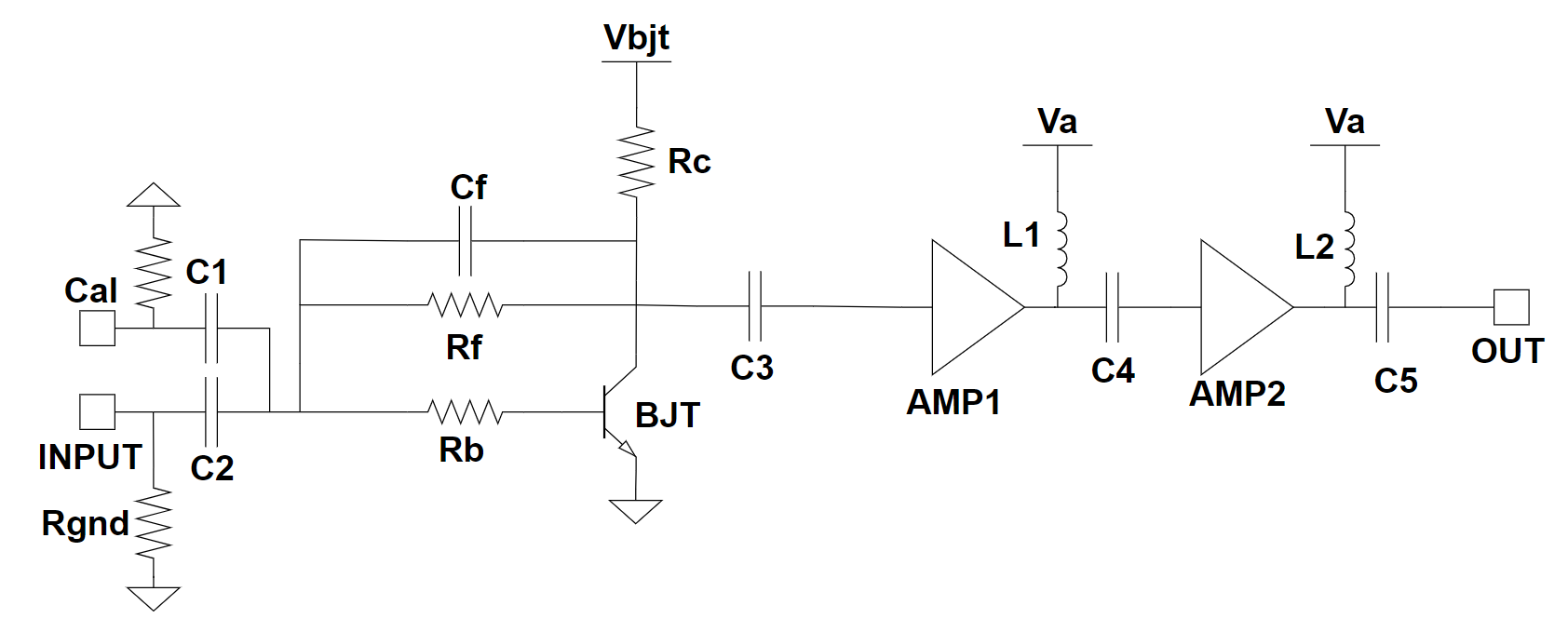}
    \caption{Simplified schematic diagram of the preamplifier board~\cite{DeZhang:2026eqi}.}
    \label{NBCircuit}
\end{figure}

Based on the preamplifier board designed to readout the LGAD~\cite{GE2022167222}, a dedicated preamplifier board was designed to read out the response signal of the ultra-thin 3D sensors generated by the external radioactive source. Fig.~\ref{NBCircuit} shows the simplified schematic diagram of the new preamplifier board. The board contains a fast Trans-impedance Amplifier (TIA) based on a Radio Frequency (RF) transistor, followed by two LTC6431-20 amplifiers, which have a gain of 20 dB in the frequency range of 100 MHz $-$ 1.2 GHz. The preamplifier board operates within a $-$3 dB bandwidth of 34.0 $-$ 594.3 MHz, and the gain of this board is approximately 116.31 mV$\cdot$ns/fC, as reported in ~\cite{DeZhang:2026eqi}. A trade-off has been made between gain and bandwidth; the current gain makes signals large enough to be recorded by a 40 GS/s digitizing oscilloscope with 4 GHz bandwidth. 

\begin{figure}[htbp]
    \centering
    \subfigure[]{\includegraphics[width=2.25in]{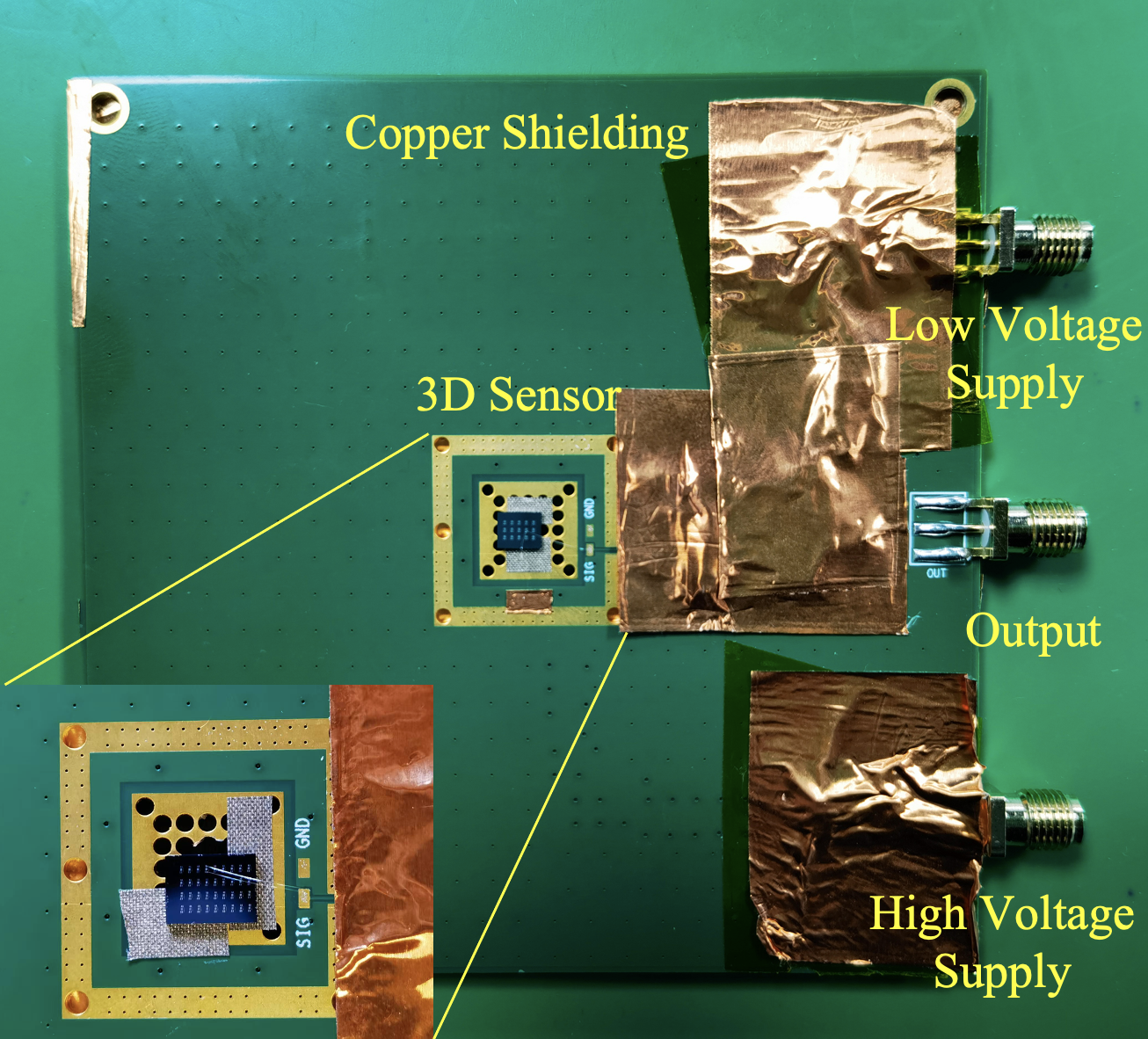}\label{PCBwithSensor}}
    \subfigure[]{\includegraphics[width=2.72in]{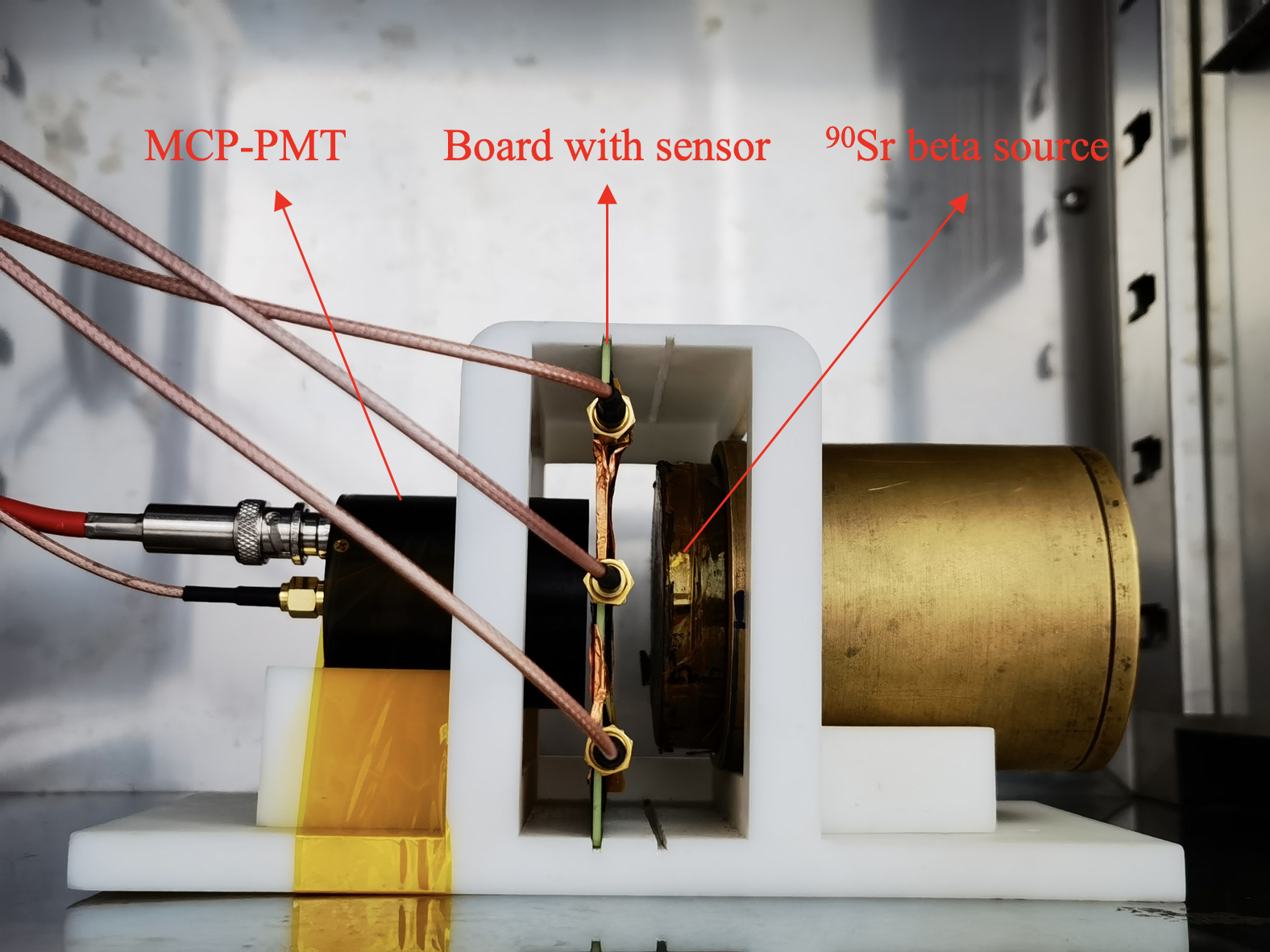}\label{Betasetup}}
    \caption{(a) The single channel preamplifier board mounted with the 3D sensor. (b) The setup of the $^{90}$Sr beta source measurements.}
    \label{Betatest}
\end{figure}

The 5 $\times$ 5 array sensor was attached to the single channel amplifier board using a double-sided conductive tape. Considering the smaller active area for the sensor, all nine pixels in the center were connected to the signal input pad to increase the coincidence rate. The board mounted with a 3D sensor is shown in Fig.~\ref{PCBwithSensor}. The electronic components on the board are covered by grounded copper tapes. The trigger was provided by the reference  Micro-Channel-Plate Photomultiplier Tube (MCP-PMT), with a time resolution of about 10 ps~\cite{Bortfeldt:2019zel}. Besides, a 50 mV threshold, corresponding to about 4.4 times the noise, was required in the 3D sensor. When testing, the MCP-PMT and the board were placed in an environmental chamber. The mechanical tools provide precise alignment of all components, as shown in Fig.~\ref{Betasetup}. All measurements were performed at \SI{20}{\celsius}. 

\section{Measurement results}
\label{Measurement results}

\subsection{Leakage-current results}

\begin{figure}[htbp]
    \centering
    \subfigure{\includegraphics[width=3.6in]{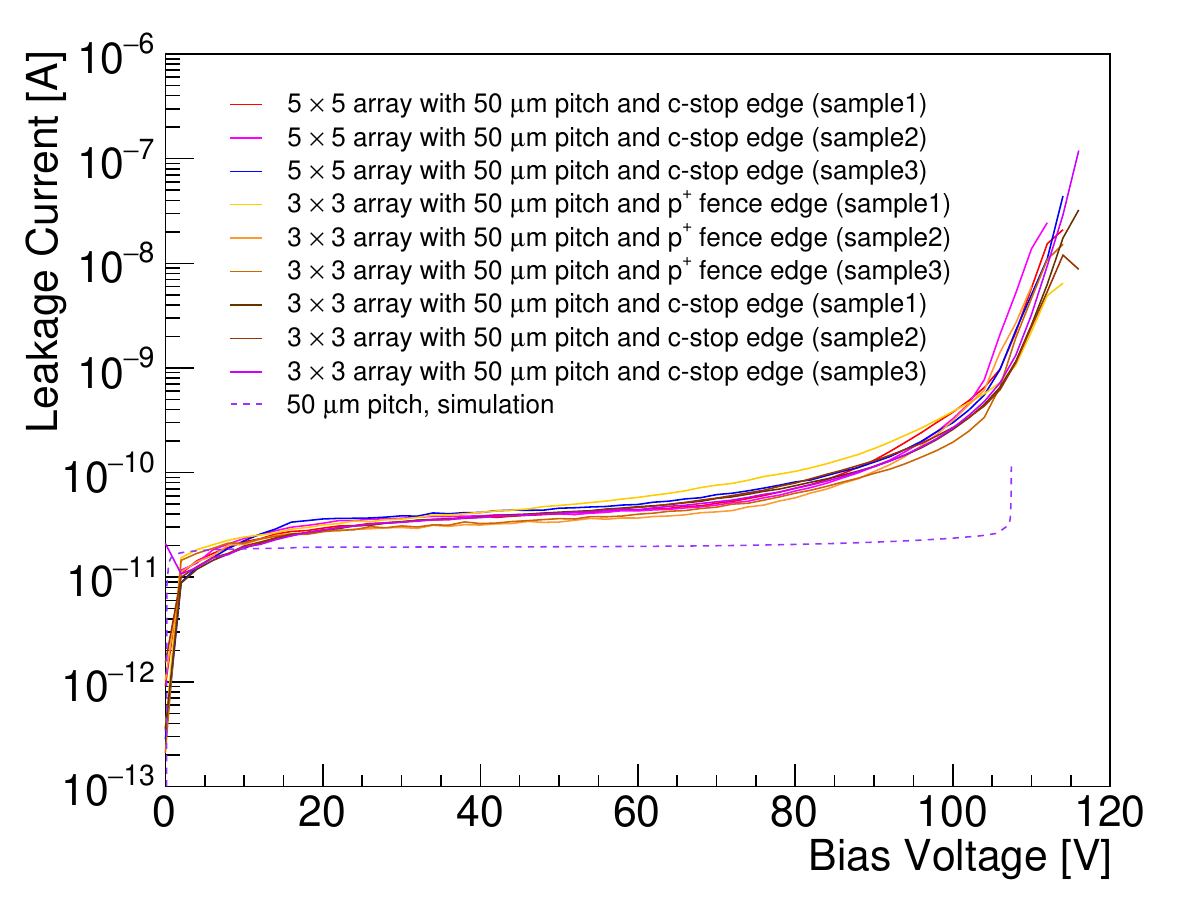}\label{3DIV_pitch50}}
    \subfigure{\includegraphics[width=3.6in]{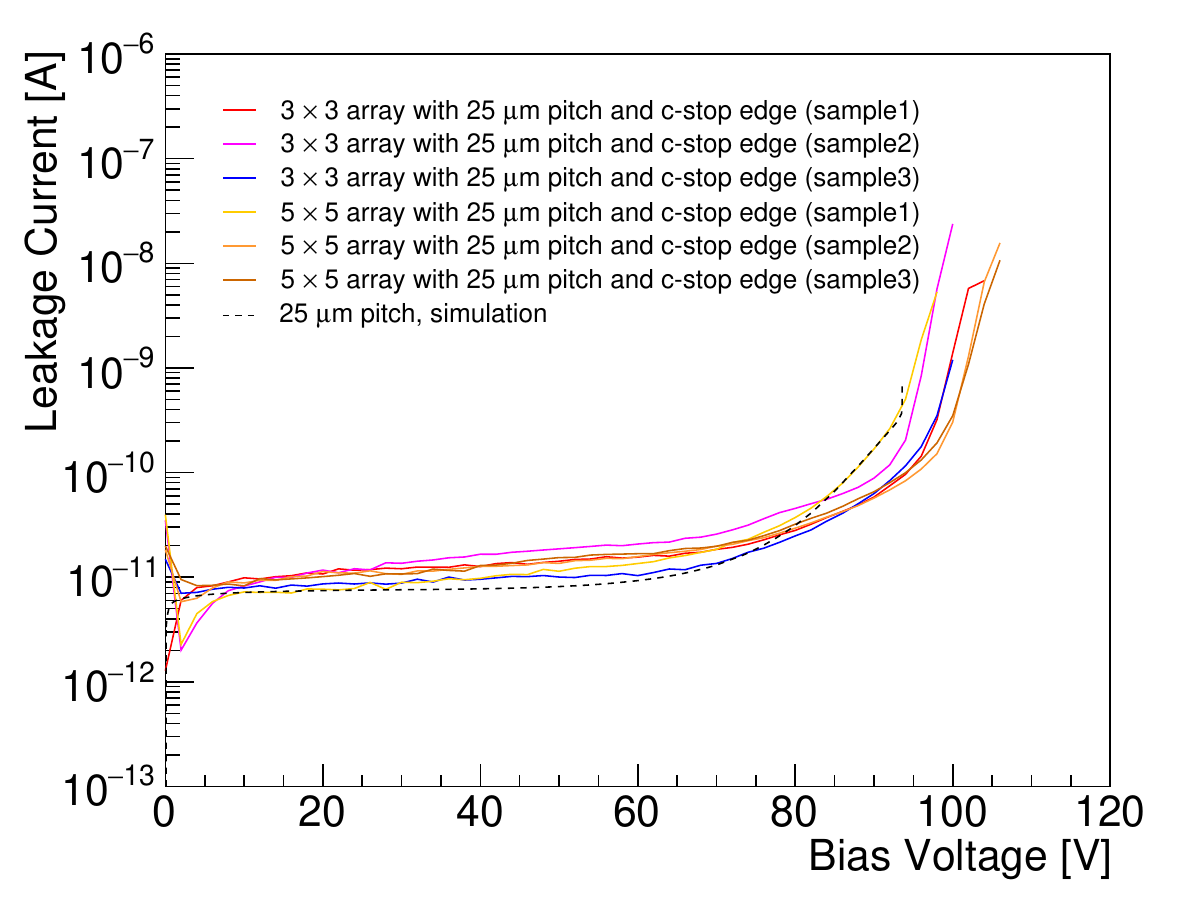}\label{3DIV_pitch25}}
    \caption{Leakage current versus bias voltage curves of a single pixel for the (a) sensors with \SI{50}{\micro m} pitch and (b) sensors with \SI{25}{\micro m} pitch, respectively. The solid lines represent the measured results, while the dashed lines represent the simulated results.}
    \label{measurediv}
\end{figure}

The measurements were performed at the wafer level on more than 3000 sensors. Fig.~\ref{measurediv} shows the central pixel's leakage current as a function of bias voltage of the representative small-array sensors. The leakage current increases at lower bias voltage, and then reaches saturation within the range of 10 $-$ 20~V at a low leakage current level. For 50 \SI{}{\micro m} pitch, the saturation current is typically in the range of 0.01 $-$ 0.1 nA, and it is in the range of 0.005 $-$ 0.05 nA for 25 \SI{}{\micro m} pitch. Finally, the breakdown occurs in the range of 100 $-$ 115~V for 50 \SI{}{\micro m} pitch, and it occurs in the range of 90 $-$ 105~V for 25 \SI{}{\micro m} pitch. The saturation current and breakdown voltage of the pixel with 50 \SI{}{\micro m} pitch is larger because of the geometry differences, which is in good agreement with the simulated results also shown in the figure. 

\subsection{Time resolution}

\begin{figure}[htbp]
    \centering
    \includegraphics[width=3.4in]{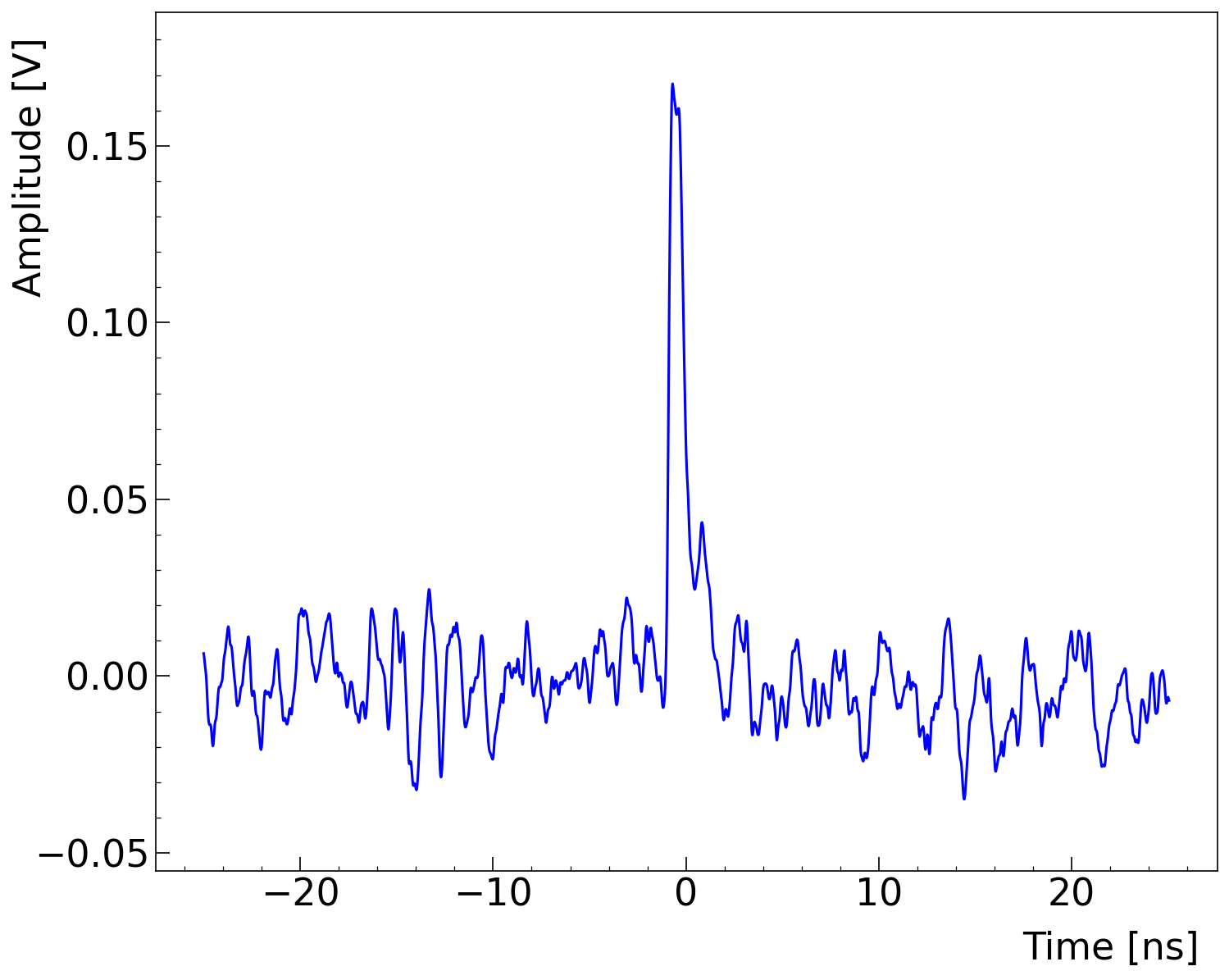}\label{P25_70V}
    \caption{A recorded waveform of the 3D sensor with 25 \SI{}{\micro m} pitch when the bias voltage is 70~V.}
    \label{waveform:P25_70V}
\end{figure}

Fig.~\ref{waveform:P25_70V} shows a recorded waveform of the 3D sensor generated by the $^{90}$Sr radioactive source. The maximum amplitude (Amax) of the signal is located in the time window of $-$5 to 5 ns, which is defined as the signal region (SR). To estimate the background, the Noise RMS is calculated as the standard deviation of the waveform in the time window of $-$25 to $-$15 ns, assuming it is unrelated to the signal and can be used to estimate the noise in the SR. The time-of-arrival (TOA) is determined as the time where the waveform exceeds half of its Amax. When calculating the TOA, the linear interpolation is applied between two adjacent samples.

\begin{figure}[htbp]
    \centering
    \subfigure[]{\includegraphics[width=2.6in]{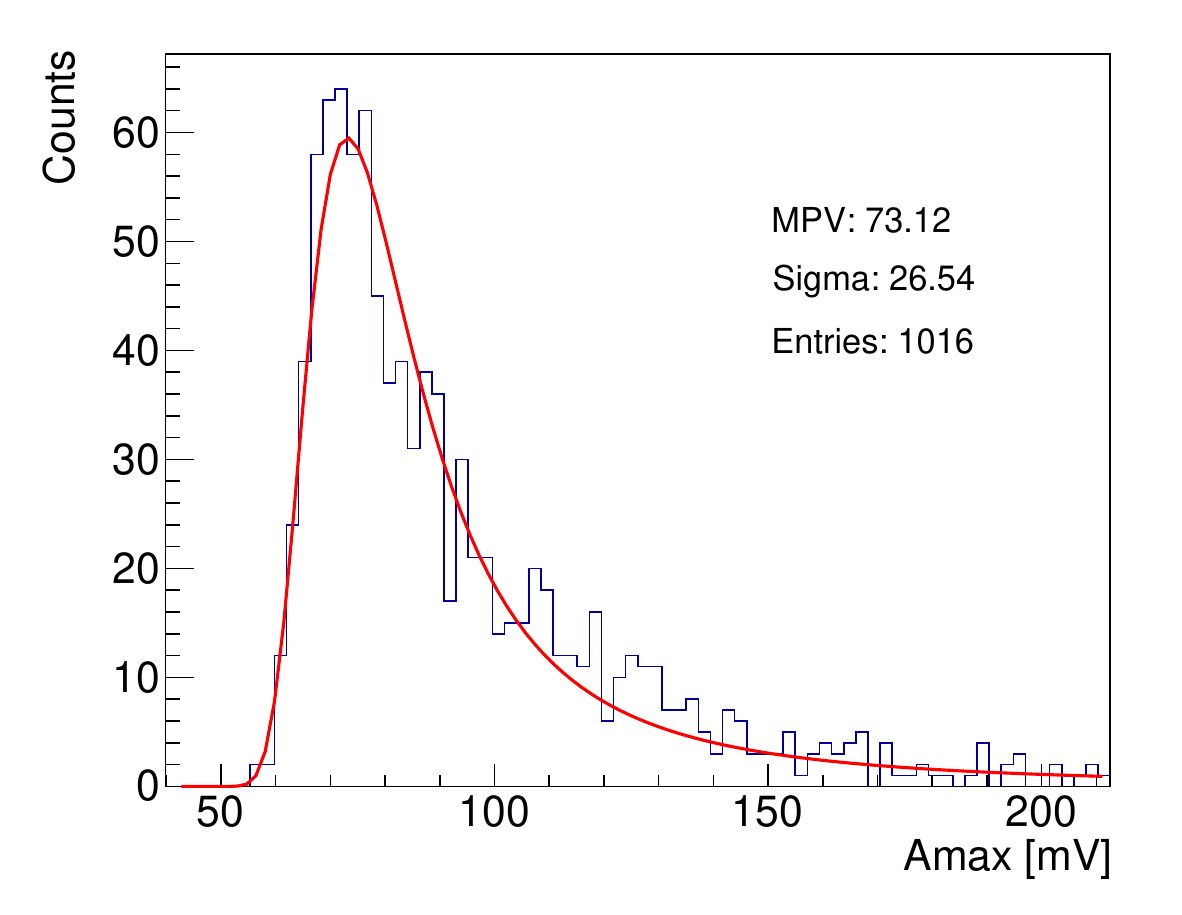}\label{P50_70V_amplitude}}
    \subfigure[]{\includegraphics[width=2.6in]{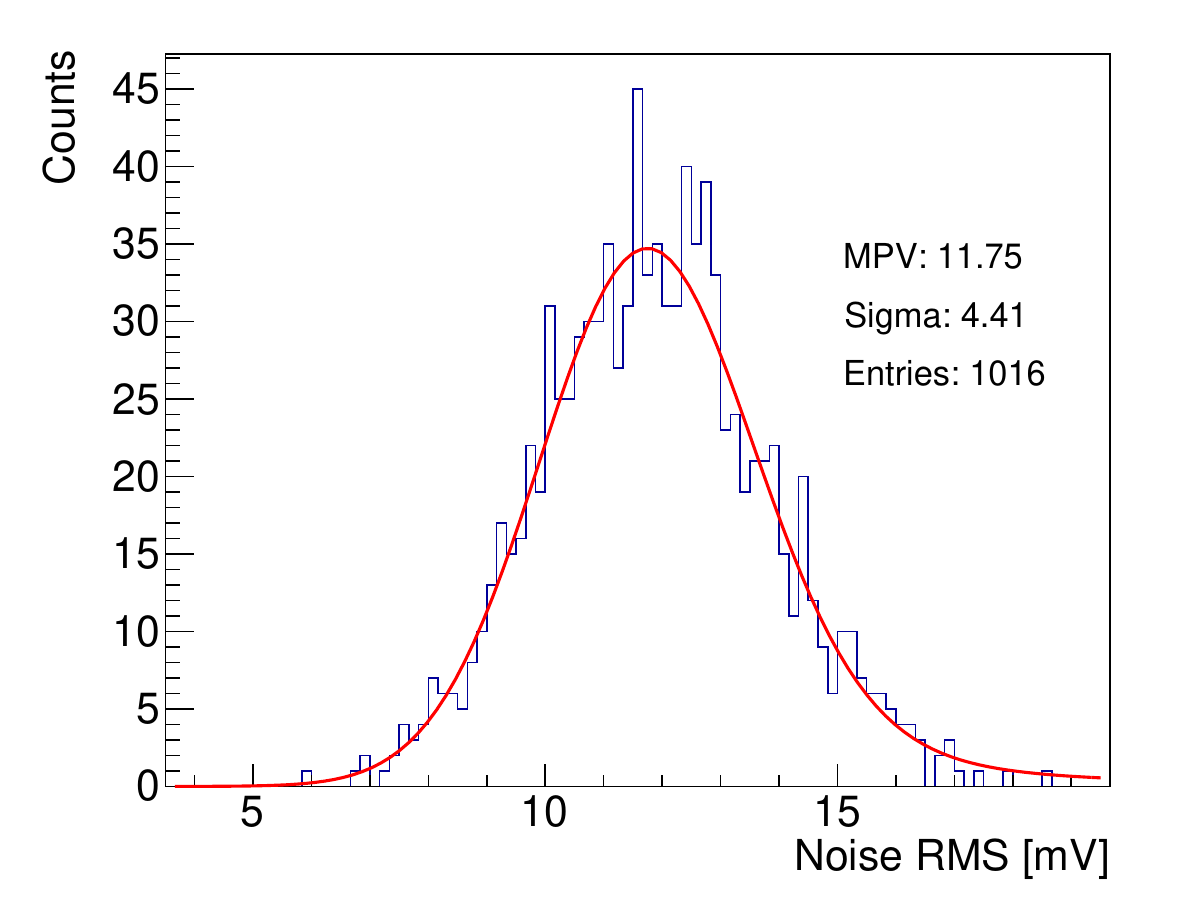}\label{P50_70V_noiserms}}
    \subfigure[]{\includegraphics[width=2.6in]{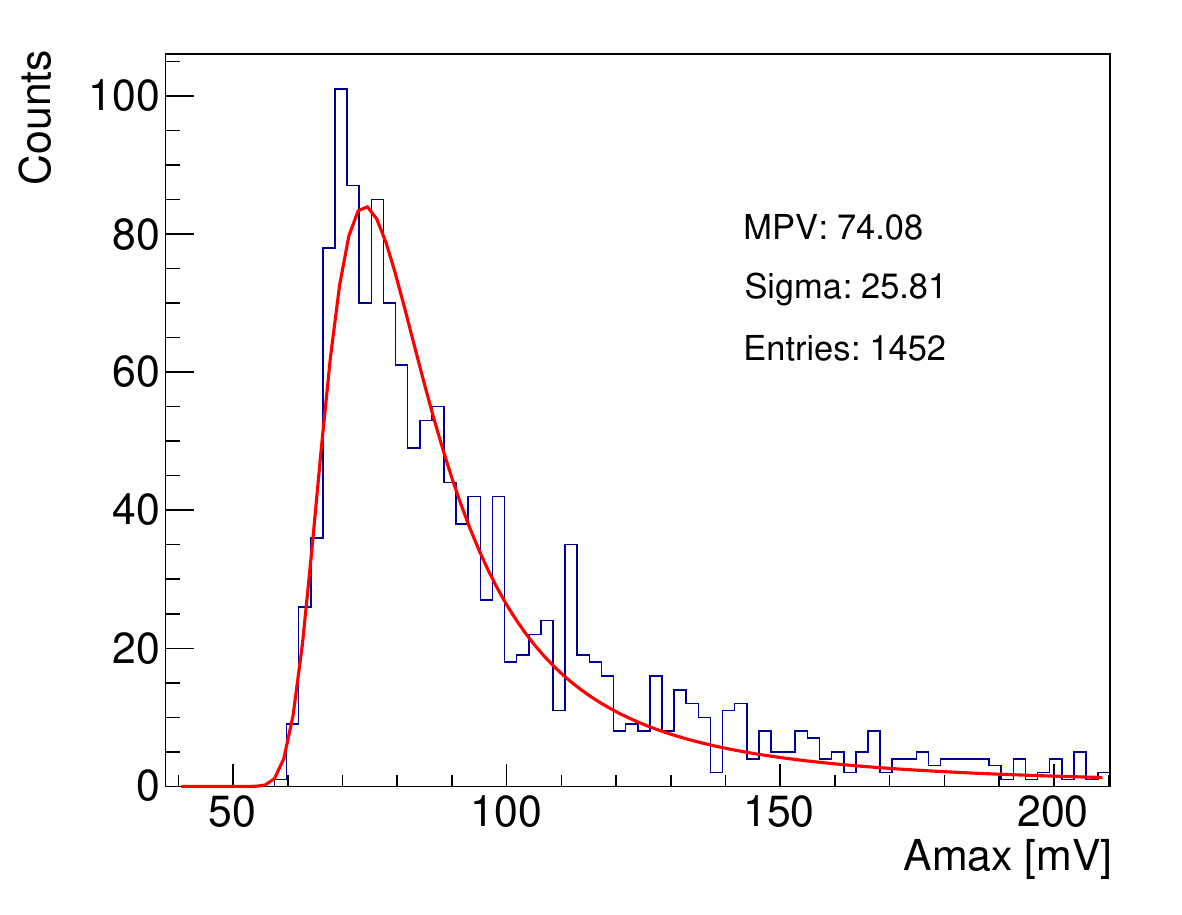}\label{P25_70V_amplitude}}
    \subfigure[]{\includegraphics[width=2.6in]{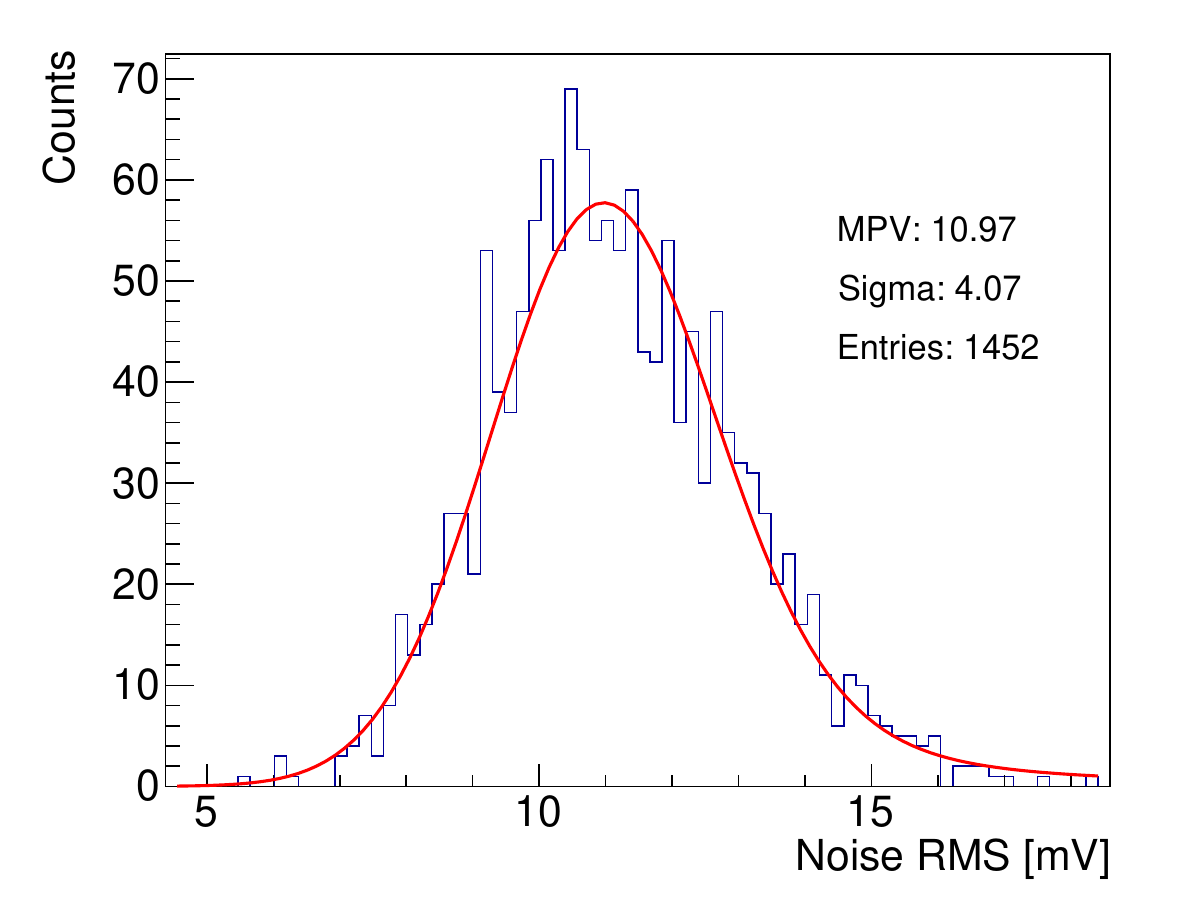}\label{P25_70V_noiserms}}
    \caption{The signal amplitude distribution (a) and the noise distribution (b) of the 3D sensor with 50 \SI{}{\micro m} pitch, operated at 110~V. The signal amplitude distribution (c) and the noise distribution (d) of the 3D sensor with 25 \SI{}{\micro m} pitch, operated at 96~V.}
    \label{sn}
\end{figure}

The signal amplitude distribution is fitted with a landau function convoluted with a gaussian function to extract the most probable value (MPV) as the amplitude at the corresponding bias voltages. And a gaussian function is used to fit the noise. Fig.~\ref{sn} shows the distributions of signal amplitude and noise from the 3D sensor with 50 \SI{}{\micro m} pitch and 25 \SI{}{\micro m} pitch. These results are obtained when the sensors operate at the highest voltages. The bias voltage is 110~V for the 50 \SI{}{\micro m} pitch sensor, and it is 96~V for the 25 \SI{}{\micro m} pitch sensor. The signal-to-noise (S/N) of these sensors is about 6$-$7.  

\begin{figure}[htbp]
    \centering
    \includegraphics[width=3.6in]{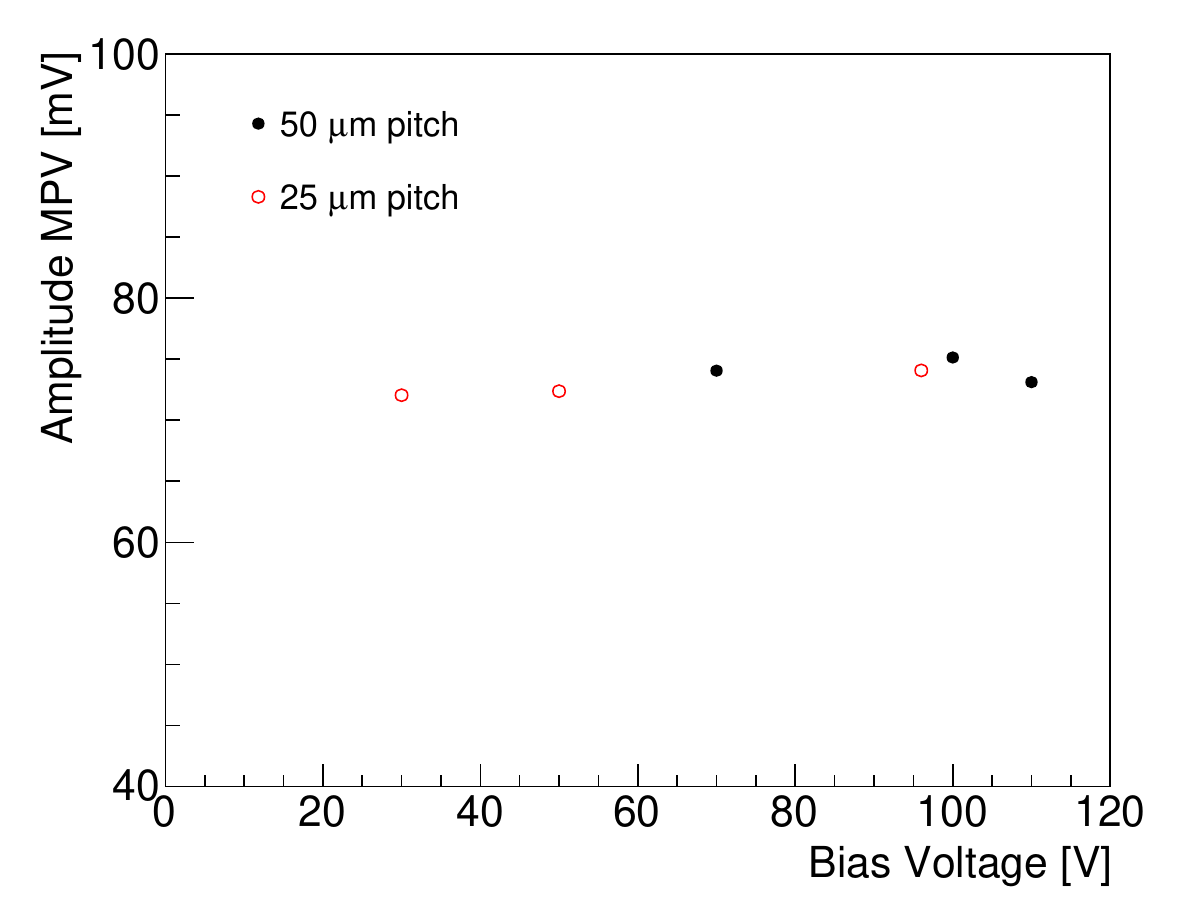}
    \caption{Amplitude MPV of 3D sensors with pitches of 50 \SI{}{\micro m} and 25 \SI{}{\micro m} as a function of the bias voltage.}
    \label{AmplitudeMPV}
\end{figure}

Fig.~\ref{AmplitudeMPV} shows the MPV of the signal amplitude as a function of bias voltage. The MPV exhibits only a small dependence, indicating that the collected charge of the 3D sensors with pitches of \SI{50}{\micro m} and \SI{25}{\micro m} remains stable over the measured bias voltage range. This behavior is consistent with the sensors being effectively fully depleted in this voltage range, as indicated by the saturation of the leakage current in the measured $I-V$ curves.

\begin{figure}[htbp]
    \centering
    \subfigure[]{\includegraphics[width=3.6in]{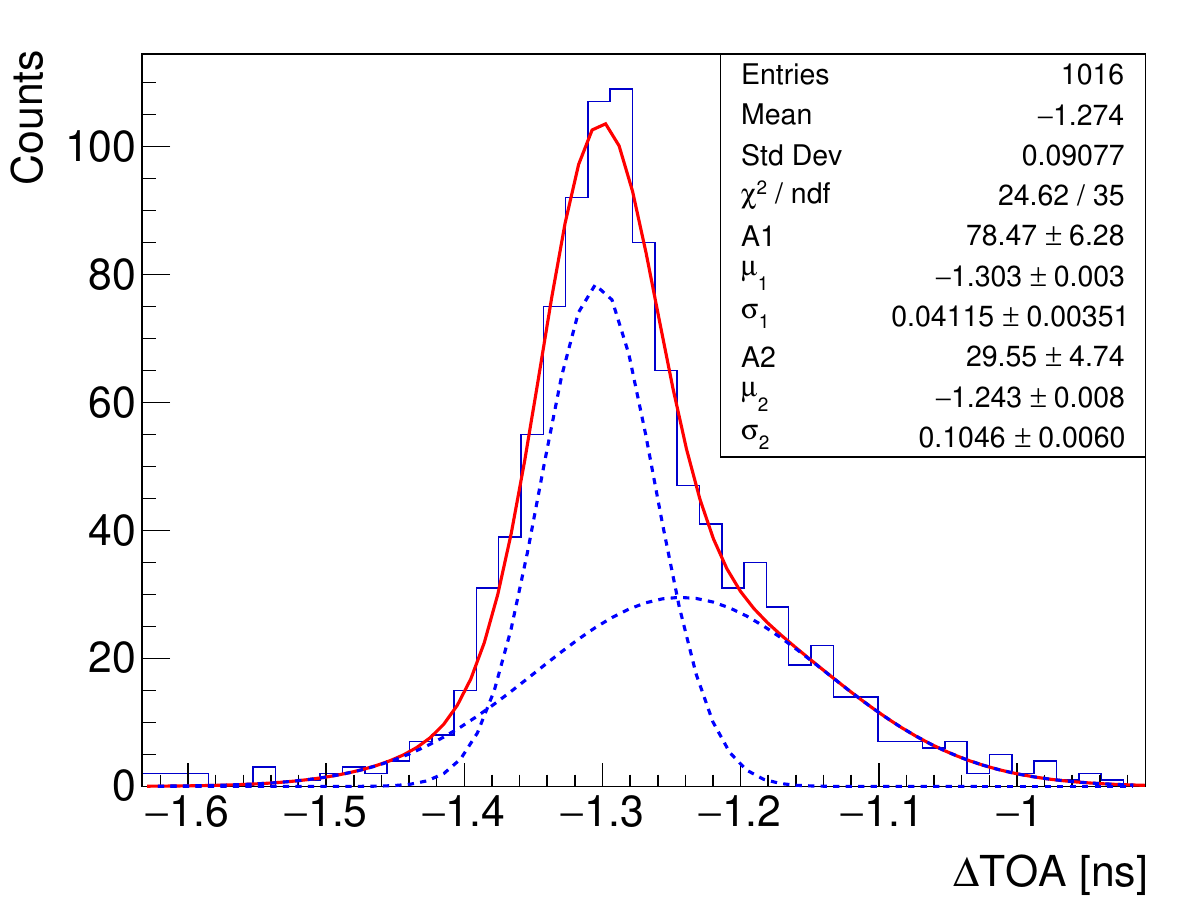}\label{P50_70V_dtoa}}
    \subfigure[]{\includegraphics[width=3.6in]{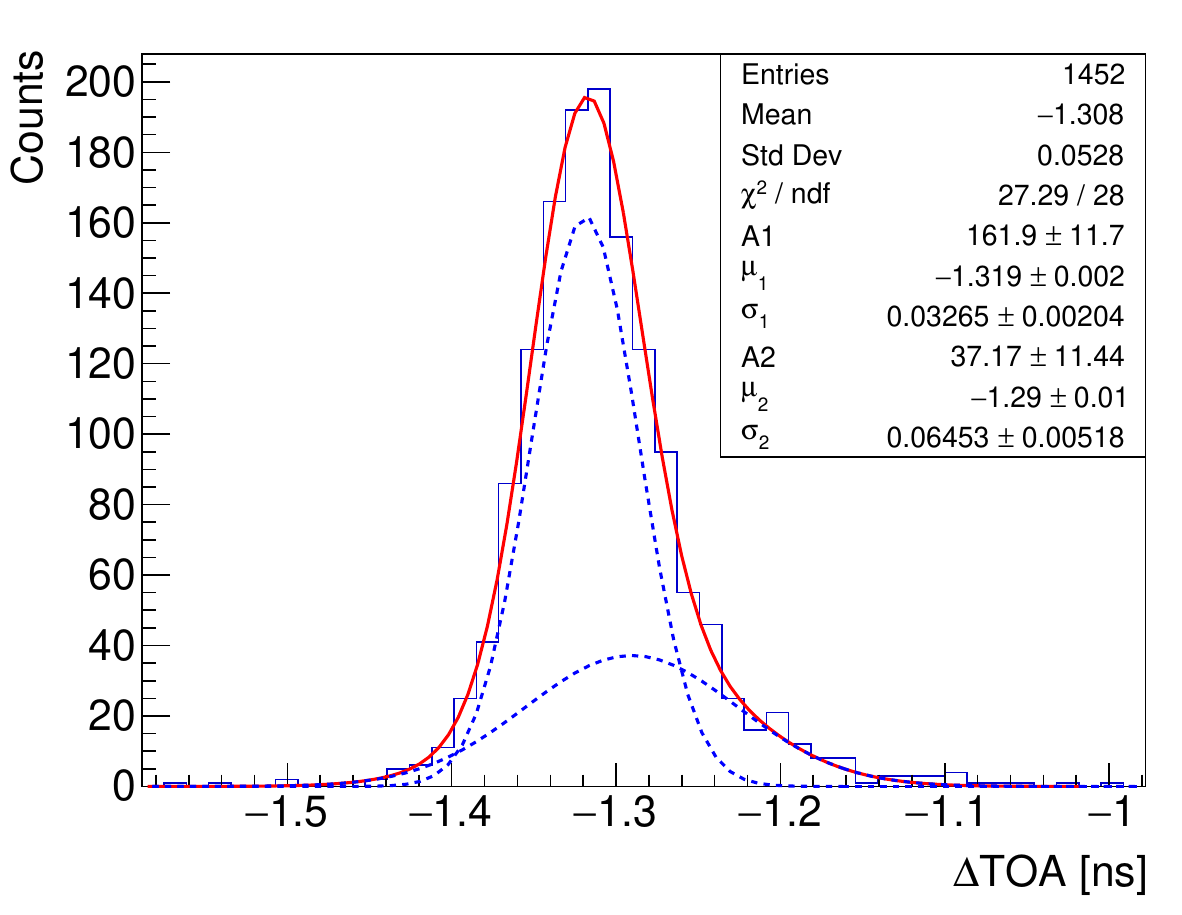}\label{P25_70V_dtoa}}
    \caption{The distribution of $\Delta$TOA between the 3D sensor and MCP-PMT: (a) The 3D sensor with 50 \SI{}{\micro m} pitch, operated at 110~V. (b) The 3D sensor with 25 \SI{}{\micro m} pitch, operated at 96~V. The distributions are fitted with the sum of two gaussian functions (blue dashed lines).}
    \label{minimum_dtoa}
\end{figure}

Fig.~\ref{minimum_dtoa} shows the distributions of TOA differences ($\Delta$TOA) between the 3D sensor and MCP-PMT, which consists of a dominating peak and a long tail. The asymmetry is mainly because the carriers generated by the incident particles in the low electric field region require a longer drift time. Therefore, the sum of two gaussian functions is applied to fit the distribution of $\Delta$TOA to extract the $\sigma_{\Delta \mathrm{TOA}}$. Here, $\sigma_1$ and $\sigma_2$ reflect the minimum and maximum values of $\sigma_{\Delta \mathrm{TOA}}$. For reference, the standard deviation (Std Dev) of the histogram is also used to evaluate the overall performance and is denoted as $\sigma_{\Delta \mathrm{TOA}}$. After subtracting the MCP-PMT time resolution in quadrature, the upper limit ($\sigma_{ \mathrm{Upper~limit}}$) and the lower limit ($\sigma_{ \mathrm{Lower~limit}}$) of the time resolution of the 3D sensor ($\mathrm{\sigma{_{3D}}}$) can be obtained, and compared to the time resolution extracted from the Std Dev ($\sigma_{ \mathrm{std~dev}}$). The uncertainty of $\sigma_\mathrm{{\Delta TOA}}$, which is the fitting error in Fig.~\ref{minimum_dtoa}, is propagated to estimate the uncertainty of $\sigma{_\mathrm{3D}}$.

\begin{figure}[htbp]
    \centering
    \includegraphics[width=3.6in]{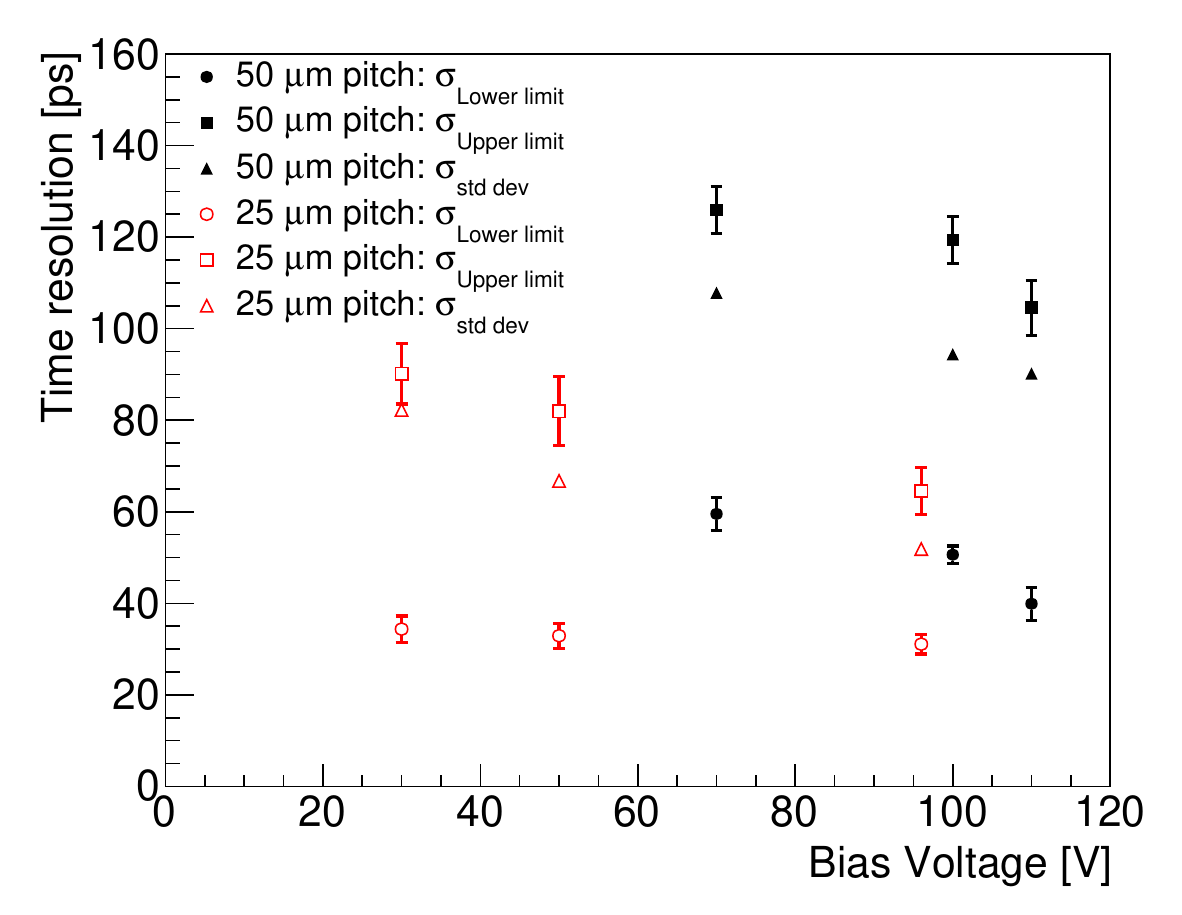}
    \caption{Time resolution of 3D sensors with pitches of 50 \SI{}{\micro m} and 25 \SI{}{\micro m} as a function of the bias voltage. The "Lower limit", "Upper Limit", and "Std Dev" of the time resolution for the 3D sensor is extracted when setting the $\sigma{_\mathrm{TOA}}$ to $\sigma_1$, $\sigma_2$ or Std Dev.}
    \label{Timemip}
\end{figure}

The time resolution as a function of bias voltage for the 3D sensor with two different pitches is summarized in Table~\ref{table:The time resolution of the 3D sensors.}. For the 3D sensor with 50 \SI{}{\micro m} pitch, the time resolution at 110~V is in the range of 39.92 $-$ 104.12 ps, while the time resolution at 96~V is in the range of 31.08 $-$ 63.75~ps for the 3D sensor with 25 \SI{}{\micro m} pitch. As shown in Fig.~\ref{Timemip}, The upper limit of time resolution decreases significantly with increasing bias voltage because the low electric field region of both sensors decreases as the voltage increases. For the sensor with a 50 \SI{}{\micro m} pitch, the lower limit of time resolution also decreases significantly with increasing bias voltage, which is related to the low electric field near the n$^{+}$ electrodes, shown in Fig.~\ref{E-Field2d_pitch50_70v}. As the bias voltage increases, the electric field in the region at the top of the n$^{+}$ electrode becomes stronger, as shown in Fig.~\ref{E-Field_pitch50_z37p5}, which enables the rapid collection of carriers~\cite{Kramberger:2019ygz}. For the geometry of 25 \SI{}{\micro m} pitch, the electric field around the whole n$^{+}$ electrodes is sufficiently strong at lower voltages, as shown in Figs.~\ref{E-Field2d_pitch25_70v},~\ref{E-Field_pitch25_z37p5}, so the reduction of the lower limit of time resolution is minimal.

\begin{table}[htbp]
\caption{Time resolution of the 3D sensors with two different pitches at different bias voltages. The "Lower limit", "Upper Limit", and "Std Dev" of the time resolution for the 3D sensor is extracted when setting the $\sigma{_\mathrm{TOA}}$ to $\sigma_1$, $\sigma_2$ or Std Dev.}
\centering
\renewcommand{\arraystretch}{1.2}
\begin{tabular}{ccccc}
    \toprule
    \multirow{2}{*}{Pitch size [\SI{}{\micro m}]} & \multirow{2}{*}{Bias voltage [V]} & \multicolumn{3}{c}{Time resolution [ps]} \\
     &  & $\sigma_{ \mathrm{Lower~limit}}$ & $\sigma_{ \mathrm{Upper~limit}}$ & $\sigma_{ \mathrm{std~dev}}$\\
    \hline
    \multirow{3}{*}{50} & 70 & 59.54 $\pm$ 3.60 & 125.50 $\pm$ 5.12 & 107.84\\
     & 100 & 50.63 $\pm$ 1.90 & 118.98 $\pm$ 5.12 & 94.41\\
     & 110 & 39.92 $\pm$ 3.62 & 104.12 $\pm$ 6.00 & 90.22\\
    \hline
    \multirow{3}{*}{25} & 30 & 34.36 $\pm$ 2.86 & 89.59 $\pm$ 6.65 & 82.25\\
     & 50 & 32.91 $\pm$ 2.72 & 81.38 $\pm$ 7.62 & 66.72\\
     & 96 & 31.08 $\pm$ 2.14 & 63.75 $\pm$ 5.24 & 51.84\\
    \bottomrule   
\end{tabular}
\label{table:The time resolution of the 3D sensors.}
\end{table}

Table~\ref{table:summaryoftiming} summarizes the time resolution of USTC-NRFC 3D pixel sensors and 3D pixel sensors from other vendors (CNM and FBK). Both the 3D sensors developed in this work and the CNM 3D sensors employ columnar electrodes. For the same pixel pitch of \SI{50}{\micro\meter}, the CNM 3D sensors achieve time resolutions of approximately 30 ps for active thicknesses ranging from \SIrange{150}{285}{\micro\meter}~\cite{Kramberger:2019ygz,Diehl:2024kli}, which is slightly better than that of the USTC-NRFC 3D sensors. Notably, the time resolution of the \SI{25}{\micro\meter}‑pitch 3D sensors is reported here for the first time, and the observed improvement compared to the \SI{50}{\micro\meter}‑pitch 3D sensors is attributed to the small cell geometry. Among all sensors compared, the FBK 3D sensors with trenched electrodes exhibit the best time resolution (approximately 12 ps)~\cite{10.3389/fphy.2024.1393019}, owing to their more uniform electric and weighting fields. 

The comparison presented here is preliminary, as the experimental conditions, such as temperature and amplifier board, differ between measurements. It is expected that time resolution of USTC-NRFC 3D pixel sensors will improve at low temperatures, due to faster carrier drift and lower noise. Further optimization of the amplifier board is necessary to reduce the electronic noise~\cite{DeZhang:2026eqi}. 

\begin{table}[htbp]
\centering
\caption{Comparison of the measured time resolution among USTC-NRFC 3D pixel sensors and 3D electrode pixel sensors from other vendors (CNM and FBK). Thickness refers to the active thickness of the sensors and R.T. is the abbreviation for room temperature.}
\begin{tabular}{cccc}
    \toprule
    Sensors & Pixel size [\SI{}{\micro m}] & Thickness [\SI{}{\micro m}] & Time resolution [ps] \\
    \midrule
    \raisebox{-0.15cm}{\multirow{2}{*}{\makecell{USTC-NRFC 3D sensors \\ (columnar electrodes)}}} & 25 $\times$ 25 & 50 & \makecell{[31, 64]\\ (96~V, \SI{20}{\celsius})}\\
     & 50 $\times$ 50 & 50 & \makecell{[40, 104]\\ (110~V, \SI{20}{\celsius})}\\
    \hline  
    \raisebox{-0.25cm}{\multirow{3}{*}{\makecell{CNM 3D sensors \\ (columnar electrodes)}}} & \raisebox{-0.25cm}{\multirow{3}{*}{50 $\times$ 50}} & 150 & \makecell{$\sim$32\\ (80~V, $-$\SI{20}{\celsius})}\\ 
     &  & 235 & \makecell{$\sim$31\\ (80~V, $-$\SI{20}{\celsius})}\\
     &  & 285 & \makecell{$\sim$30\\ (100~V, $-$\SI{20}{\celsius})}\\
    \hline
    \raisebox{-0.25cm}{\makecell{FBK 3D sensors \\ (trenched electrodes)}} & \raisebox{-0.25cm}{55 $\times$ 55} & \raisebox{-0.25cm}{150} & \raisebox{-0.25cm}{\makecell{$\sim$12\\ (100~V, R.T.)}}\\
    \bottomrule 
\end{tabular}
\label{table:summaryoftiming}
\end{table}

\section{Conclusion and outlook}
\label{conclusion}

3D silicon sensors with ultra-thin active substrates (50 \SI{}{\micro m}) and small-pitch (50 \SI{}{\micro m} and 25 \SI{}{\micro m}) were successfully fabricated in the first prototype run at USTC NRFC. The fabrication process was developed based on the single-sided process, among which it is worth mentioning the use of a modified procedure of Bosch process to etch high-quality columns, as well as the use of solid source wafers to introduce impurity atoms into etched columns. The $I-V$ results show good characteristics of test structures with leakage currents less than 0.1 nA per pixel, and with breakdown voltage larger than 90 V, which are in good agreement with the simulated results. $^{90}$Sr source measurements results show that the time resolution is in the range of 39.92 $-$ 104.12~ps at 110~V and \SI{20}{\degree} for the 3D sensor with 50 \SI{}{\micro m} pitch, and it is in the range of 31.08 $-$ 63.75~ps at 96~V and \SI{20}{\degree} for the 3D sensor with 25 \SI{}{\micro m} pitch. 

To achieve better time resolution, both the sensor and the electronics need further optimization. Optimizing the non-uniform electric field distribution of the sensor is very crucial. At the same time, the breakdown voltage of the sensor also needs to be improved. Moreover, since the signal generated by MIP particles in an ultra-thin substrate is very small, increasing the substrate thickness of the sensor will also be considered in the future. Regarding the electronics, in addition to increasing the bandwidth of the preamplifier board, optimization is also needed to reduce the noise coming from the board.

\section*{Acknowledgment}
This work was partially carried out at the USTC Center for Micro and Nanoscale Research and Fabrication. We are grateful for all staff who participated in this work.

\bibliographystyle{elsarticle-num-names}
\bibliography{reference}

\end{document}